\documentclass[12pt]{article}
\usepackage{amsfonts,amsmath,eucal,amssymb,latexsym,amsbsy}

\usepackage[dvips]{graphicx}

\def \beq {\begin{eqnarray}}
\def \eeq {\end{eqnarray}}
\def \beqn {\begin{eqnarray*}}
\def \eeqn {\end{eqnarray*}}

\newcommand{\halmos}{\rule{1ex}{1.4ex}}

\newcounter{for}[section]

\newtheorem{theorem}{Theorem}
\newtheorem{itlemma}{Lemma}
\newtheorem{itproposition}{Proposition}
\newtheorem{itcorollary}{Corollary}
\newtheorem{itremark}{Remark}
\newtheorem{itremarks}{Remarks}
\newtheorem{itdefinition}{Definition}
\newtheorem{itexample}{Example}
\newtheorem{itclaim}{Claim}
\newtheorem{itfact}{Fact}

\newenvironment{fact}{\begin{itfact}\rm}{\end{itfact}}
\newenvironment{claim}{\begin{itclaim}\rm}{\end{itclaim}}
\newenvironment{lemma}{\begin{itlemma}}{\end{itlemma}}
\newenvironment{remark}{\begin{itremark}\rm}{\end{itremark}}
\newenvironment{remarks}{\begin{itremarks} \rm}{\end{itremarks}}
\newenvironment{corollary}{\begin{itcorollary}}{\end{itcorollary}}
\newenvironment{proposition}{\begin{itproposition}}{\end{itproposition}}
\newenvironment{definition}{\begin{itdefinition}\rm}{\end{itdefinition}}
\newenvironment{example}{\begin{itexample}\rm}{\end{itexample}}
\newenvironment{proof}{\noindent {\em Proof}.\ \
}{\hspace*{\fill}$\halmos$\medskip}
\newcommand{\be}[1]{\addtocounter{for}{1} \begin{equation}\label{#1}}
\newcommand{\ee}{\end{equation}}
\newcommand{\bl}[1]{\begin{lemma}\label{#1}.}
\newcommand{\br}[1]{\begin{remark}\label{#1}.}
\newcommand{\brs}[1]{\begin{remarks}\label{#1}.}
\newcommand{\bt}[1]{\begin{theorem}\label{#1}.}
\newcommand{\bd}[1]{\begin{definition}\label{#1}.}
\newcommand{\bp}[1]{\begin{proposition}\label{#1}.}
\newcommand{\bc}[1]{\begin{corollary}\label{#1}.}
\newcommand{\bfact}[1]{\begin{fact}\label{#1}.}
\newcommand{\bex}[1]{\begin{example}\label{#1}.}
\newcommand{\ec}{\end{corollary}}
\newcommand{\efact}{\end{fact}}
\newcommand{\eex}{\end{example}}
\newcommand{\el}{\end{lemma}}
\newcommand{\er}{\end{remark}}
\newcommand{\ers}{\end{remarks}}
\newcommand{\et}{\end{theorem}}
\newcommand{\ed}{\end{definition}}
\newcommand{\ep}{\end{proposition}}
\newcommand{\epr}{\end{proof}}
\newcommand{\bpr}{\begin{proof}}
\newcommand{\bcl}[1]{\begin{claim}\label{#1}}
\newcommand{\ecl}{\end{claim}}

\newcommand{\ecs}{\end{corollary}}
\newcommand{\eers}{\end{exercise}}
\newcommand{\eexs}{\end{example}}
\newcommand{\eems}{\end{example}}
\newcommand{\els}{\end{lemma}}
\newcommand{\eles}{\end{lemmaex}}
\newcommand{\ets}{\end{theorem}}
\newcommand{\eds}{\end{definition}}
\newcommand{\eps}{\end{proposition}}

\newcommand{\bi}{\begin{itemize}}
\newcommand{\ei}{\end{itemize}}
\newcommand{\ben}{\begin{enumerate}}
\newcommand{\een}{\end{enumerate}}

\def\vbar{\mathchoice{\vrule height6.3ptdepth-.5ptwidth.8pt\kern-.8pt}
   {\vrule height6.3ptdepth-.5ptwidth.8pt\kern-.8pt}
   {\vrule height4.1ptdepth-.35ptwidth.6pt\kern-.6pt}
   {\vrule height3.1ptdepth-.25ptwidth.5pt\kern-.5pt}}
\def\fudge{\mathchoice{}{}{\mkern.5mu}{\mkern.8mu}}
\def\bbc#1#2{{\rm \mkern#2mu\vbar\mkern-#2mu#1}}
\def\bbb#1{{\rm I\mkern-3.5mu #1}}
\def\bba#1#2{{\rm #1\mkern-#2mu\fudge #1}}
\def\bb#1{{\count4=`#1 \advance\count4by-64 \ifcase\count4\or\bba A{11.5}\or
   \bbb B\or\bbc C{5}\or\bbb D\or\bbb E\or\bbb F \or\bbc G{5}\or\bbb H\or
   \bbb I\or\bbc J{3}\or\bbb K\or\bbb L \or\bbb M\or\bbb N\or\bbc O{5} \or
   \bbb P\or\bbc Q{5}\or\bbb R\or\bbc S{4.2}\or\bba T{10.5}\or\bbc U{5}\or
   \bba V{12}\or\bba W{16.5}\or\bba X{11}\or\bba Y{11.7}\or\bba Z{7.5}\fi}}

\def \R {{\mathbb R}}

\def \N {{\mathbb N}}
\def \P {{\mathbb P}}

\def \ra {\rightarrow }

\def \s {y}
\def \LL {{\cal{L}}}
\def \M {{\cal{M}}}

\def \PP {{\cal{P}}}

\def \D{{\cal{D}}}

\def \ind {{\bf 1}}

\def\Proof{{\sl Proof.}\quad}

\def\Proof{{\sl Proof.}\quad}
\newcommand{\fine}{\hspace*{\fill}$\halmos$\medskip}
\newcommand{\ba}[1]{\addtocounter{for}{1} \begin{eqnarray}\label{#1}}
\newcommand{\ea}{\end{eqnarray}}

\def\sqr#1#2{{\vcenter{\vbox{\hrule height .#2pt
                             \hbox{\vrule width .#2pt height#1pt \kern#1pt
                                   \vrule width .#2pt}
                             \hrule height .#2pt}}}}

\def\pmb#1{\setbox0=\hbox{#1}%
   \kern-.025em\copy0\kern-\wd0
   \kern.05em\copy0\kern-\wd0
   \kern-.025em\raise.0433em\box0 }
\def\sqr#1#2{{\vcenter{\vbox{\hrule height.#2pt
     \hbox{\vrule width.#2pt height#1pt \kern#1pt
   \vrule width.#2pt}\hrule height.#2pt}}}}

\def\i{I\!\!\!I\,}

\def\e{\epsilon}

\def\e{\epsilon}
\def\d{\delta}
\def\l{\lambda}

\def\g{\gamma}

\def\a{\alpha}
\def\b{\beta}
\def\r{\rho}

\def\ua{\underline{\a}}
\def\ub{\underline{\b}}
\def\ug{\underline{\g}}
\def\us{\underline{\s}}
\def\ul{\underline{\l}}

\begin{document}

\title{Heterogeneous credit portfolios and the dynamics of  the aggregate losses}

\author{Paolo Dai Pra
\thanks{e-mail: \texttt{daipra@math.unipd.it}}
\\[0.2cm]Dipartimento di Matematica Pura ed Applicata, Universit\`a di  Padova \\
63, Via Trieste;  I - 35121 - Padova, Italy
 \\[0.2cm]and\\[0.2cm]
Marco Tolotti\footnote{Corresponding author.  
\texttt{tolotti@unibocconi.it}, phone number: (+39) 0258365485 }\\[0.2cm]
Dipartimento di Finanza, Universit\`a Bocconi \\
25, Via Sarfatti ; I - 20136 Milano, Italy 
 }
\date{\today}
\maketitle {\abstract{We study the impact of contagion in a network
of firms facing credit risk. We describe an intensity based model
where the homogeneity assumption is broken by introducing a
\emph{random environment} that makes it possible to take into
account  the idiosyncratic characteristics of the firms. We shall
see that our model goes behind the identification of groups of firms
that can be considered basically exchangeable. Despite this
heterogeneity assumption our model has the advantage of being
totally tractable. The aim is to quantify the losses that a bank may
suffer in a large credit portfolio. Relying on a large deviation
principle on the trajectory space of the process, we state a
suitable law of large number and a central limit theorem useful to
study large portfolio losses. Simulation results are provided as
well as applications to portfolio loss
distribution analysis.\\

\vspace{0.6cm}



}}

\newpage
\section*{Introduction} During the last years the
challenging issue of describing the \emph{dynamics} of the loss
process connected with portfolios of many obligors has received more
and more attention. Applications can be found both for management
purposes (see \cite{DDDD}) and in the literature dealing with
pricing and hedging of risky derivatives as CDOs (Collateralized
Debt Obligations). For a discussion of this framework see \cite{GG}
and \cite{S}.

When dealing with \emph{portfolio} losses, it becomes crucial the
modeling of the dependence structure among the   obligors. One
standard procedure is to directly  specify  the \emph{intensity of
default} of the single obligors belonging to the portfolio in order
to infer the dynamics of the global system and thus the distribution
of the aggregate losses.  In the context of reduced form models  a
rather general framework is the conditionally Markov modeling
approach by Frey and Backhaus (see \cite{FB2} and \cite{FB}).  One
drawback of the intensity based models is the difficulty in managing
large heterogeneous portfolios because of the presence of many
obligors with different specifications. In this case it is common
practice to assume homogeneity assumptions in order to reduce the
complexity of the problem. A typical approach is to divide the
portfolio into groups where the obligors may be considered
exchangeable.

In this paper we describe an intensity based model    where the
homogeneity assumption is broken by introducing a \emph{random
environment} that makes it possible to take into account  the
idiosyncratic characteristics of the firms. We shall see that our
model goes behind the identification of groups of firms that can be
considered basically exchangeable. Despite this heterogeneity
assumption our model has the advantage of being totally tractable.

The goal is to describe the evolution of the  losses for a large
portfolio where  \emph{heterogeneity} and \emph{direct contagion}
among the firms are taken into account. We denote by $L^N(t)$ the
random variable describing the  losses at time $t\in[0,T]$ for a
portfolio of size $N$. Our approach works as follows. First we study
the $N\to\infty$ limiting distributions on the path space of some
aggregate variables useful to characterize the evolution of
$L^{N}(t)$ for $t\in[0,T]$. To this effect we shall derive an
appropriate \emph{Law of Large Numbers} based on a Large Deviations
Principle in order to describe a limiting behavior that can be
considered as a asymptotic regime with infinite firms. Finally, we
study the \emph{finite volume approximations} (for finite but large
$N$) of the limiting distribution via a suitable version of the
\emph{Central Limit Theorem} that describes the fluctuations around
this limit. In most cases, these dynamical fluctuation theorems are
proved by method of weak convergence of processes; this approach has
been widely applied to models close in spirit to this work. We quote
\cite{FB2} and \cite{DRST} for applications
 to finance. The effectiveness of those methods for heterogeneous
models is, however, unclear. We follow here a different approach,
which allows to prove a Central Limit Theorem directly in the
underlying trajectory space. This approach is based on a general
Central Limit Theorem in \cite{B}. Although various applications of
this theorem to fluctuations of Markov processes can be found in the
literature (see e.g. \cite{BB} and \cite{DDD}), to our knowledge the
first application to a non-reversible Markov process.\\

In the risk management context, our model may be useful for the
management of large portfolios, in the spirit of other models
proposed in \cite{GW} or in    \cite{DRST}. It has been remarked
that in many real world applications default are rather rare events,
so that,
 for instance, the fraction of defaulted firms is close to zero and a normal
 approximations is not meaningful. Our models and results are only concerned
 with time scales for which a proper fraction of the portfolio is likely to be affected by the defaults.

We believe that our paper may be considered as an original
contribution in the modeling of portfolio losses dynamics that
accounts for both heterogeneity and contagion. On the other hand, to
our knowledge, this is the first attempt to apply large deviations
and normal fluctuation theory on path spaces (that is, in a dynamic
fashion) for finance or credit management purposes, except for what
contained in\cite{DRST}. For a survey on large deviations methods
applied to finance and credit risk see \cite{P}.

The models we propose in this paper are the simplest heterogeneous models describing systems comprised
 by many defaultable components, whose defaults are positively correlated
 via an interaction of mean-field type, i.e. with no geometric structure.
 Although we have been inspired by financial applications, we believe the basic
  principles should apply to other contexts.

The outline of this paper is as follows. In Section 1 we
illustrate the model and the main theorems. In Section 2 we apply
these results to the  large portfolio losses analysis. Some
examples with explicit computations and simulations are also
provided. In Section 3 we conclude. Appendix A is devoted to the
proofs of the three main theorems stated in Section 1.

\section{Model and main results}

Consider a network of $N$ defaultable firms, whose states are
denoted by $\s_1,\s_2,\ldots,\s_N$, $\s_i \in \{0,1\}$. The event
$\{\s_i = 1\}$ means that the $i$-th firm has defaulted. The values
$\s_1,\s_2,\ldots,\s_N$ give rise to an {\em aggregate variable}
$m_N$ which indicates the global state of the network:
\[
m_N := \frac{1}{N} \sum_{i=1}^N \a_i \s_i,
\]
where $\a_1,\a_2,\ldots,\a_N$ are given nonnegative numbers. $\a_i$
can be interpreted as the impact the default of the $i$-th firm has
on the aggregate variable $m_N$. In order to model contagion, we
assume the instantaneous rate of default  of the
$i$-th firm is an increasing function of $m_N$. More specifically,
we assume the rate of default of the $i$-th firm  is given by
\[
\i_{\{\s_i = 0\}} e^{\b_i m_N - \g_i},
\]
where $\i_A$ is the indicator function of the set $A$, and $\b_i
\geq 0$, $\g_i \in \R$ are given constant. $\b_i$ represents the
sensibility of the $i$-th firm to variations of the aggregate
variable $m_N$, while $\g_i$ can be interpreted as the
``robustness'' of the $i$-th firm: a large value of $\g_i$ means
that the $i$-th firm is very unlikely to default within a given
time.

\noindent Thus, for any {\em fixed} value of $\ua :=
(\a_1,\a_2,\ldots,\a_N)$, $\ub := (\b_1,\b_2,\ldots,\b_N)$ and $\ug
:= (\g_1,\g_2,\ldots,\g_N)$, the variable $\us :=
(\s_1,\s_2,\ldots,\s_N)$ evolves as a Markov chain in continuous
time, with infinitesimal generator given by \be{generator}
\LL_{\ua,\ub,\ug} f(\us) := \sum_{i=1}^N \i_{\{\s_i = 0\}} e^{\b_i
m_N - \g_i} [f(\us^i) - f(\us)], \ee where $\us^i$ denotes the {\em
configuration} obtained from $\us$ by changing $\s_i$ from $0$ to
$1$. We assume the system start at time $t=0$ from the configuration
$\us(0) = (0,0,\ldots,0)$. The evolution randomly drives the network
towards the trap state $(1,1\ldots,1)$, which is reached in finite
time.

From now on we denote by $\l_i := (\a_i,\b_i,\g_i)$ the triple of
the parameters corresponding to the $i$-th firm. The $\l_i$'s model
the heterogeneity of the system. We consider here the point of view
of {\em disordered models}, i.e. we assume $\l_1,\l_2,\ldots,\l_N$
to be i.i.d. random variables, with a given law $\mu$. In order to
avoid inessential difficulties, the law $\mu$ is assumed to have
compact support in $\R^+ \times \R^+ \times \R$. Note that, for a
given $i$, the random variables $\a_i,\b_i,\g_i$ are {\em not}
assumed to be independent. Sometimes, the vector $\ul :=
(\l_1,\l_2,\ldots,\l_N)$ will be referred to as {\em random
environment}.

\noindent Consider a time $T>0$, and denote by $\us[0,T] =
(\us(t))_{t \in [0,T]}$ the trajectory described by the
configuration under the stochastic evolution (\ref{generator}). Each
component $\s_i[0,T]$ is either identically $0$ or it flips from $0$
to $1$ at the {\em default time} \be{deft} \tau_i := \inf\{t>0 :
\s_i(t) = 1\}. \ee By convention, we set $\s_i(\tau_i) = 1$. This
set of $\{0,1\}$-valued trajectories is denoted by $\D[0,T]$. Each
trajectory in $\D[0,T]$ can be identified with its default time
(which is set to be equal to $T$ if there is no default); thus
$\D[0,T]$  inherits the topology induced by the usual topology on
$\R$ for the default time. Equivalently, the topology on $\D[0,T]$
is the one induced by the Skorohod topology on the set of
$\R$-valued functions which are right-continuous and admit limit
from the left at any point of $[0,T]$ (see e.g. \cite{EK}).

\noindent In this paper we are interested in the asymptotic
behavior, as $N \ra +\infty$, of empirical averages of the form
\[
\frac{1}{N} \sum_{i=1}^N f(\s_i [0,T]) =: \int f d\rho_N (\us[0,T]),
\]
where $f:\D[0,T] \ra \R$ is a Borel measurable function, and
\[
\rho_N(\us[0,T]) := \frac{1}{N} \sum_{i=1}^N \d_{\s_i [0,T]}
\]
is called {\em empirical measure}. More generally, we shall consider
the empirical measure
\[
\rho_N(\us[0,T],\ul) := \frac{1}{N} \sum_{i=1}^N \d_{\s_i
[0,T],\l_i},
\]
which is a random measure on $\D[0,T] \times \R^3$. Note that
$\rho_N(\us[0,T])$ is the marginal of $\rho_N(\us[0,T],\ul)$ on
$\D[0,T]$.

\noindent In what follows, we denote by $\M_1$ the set of
probability measures on $\D[0,T] \times Supp(\mu)$, while $\M$
will denote the set of signed measure on $\D[0,T] \times \R^3$.
Both sets are provided with the weak topology.

\noindent For $\s[0,T] \in \D[0,T]$ with $\s(T)=1$, we set
\[
\tau(\s[0,T]) := \inf \{t > 0 : \s(t) =1\}.
\]
For $Q \in \M$ we define
\begin{multline}
\label{F} F(Q) := \int Q(d\s[0.T],d\l) \left\{ \int_0^T (1-\s(t))
\left( 1- e^{-\g} e^{\b \int Q(d\eta[0,T],d\l') \a' \eta(t)} \right)
dt \right.
\\ + \left. \s(T) \left[ - \g + \left(\b \int Q(d\eta[0,T],d\l') \a' \eta(t^-)\right) \Big|_{t=\tau(\s[0,T])} \right]
\right\},
\end{multline}
where $\l = (\a,\b,\g)$ and $\l' = (\a',\b',\g')$. For a fixed
$\ul$, the infinitesimal generator (\ref{generator}), together with
the initial condition $\us(0) = (0,0,\ldots,0)$ induces a
probability $P_N^{\ul}$ on $\D^N[0,T]$. We think of $P_N^{\ul}$ as
the {\em conditional law} of the process given the random
environment. We denote by
\[
P_N(d\us[0,T],d\ul) := P_N^{\ul}(d\us[0,T]) \otimes \mu^{\otimes
N}(d\ul)
\]
the joint law of the process and the environment. The distribution
of $\rho_N(\us[0,T],\ul)$ under $P_N$ will be denoted by $P_N \circ
\rho_N^{-1}$.

\noindent A special case is when all components of $\ul$ are zero.
In this case each firm defaults with rate $1$, independently of the
others. We denote by $W$ the law on $\D[0,T]$ of this process.

\noindent In what follows, for $Q_1,Q_2 \in \M_1$, we denote by
\[
H(Q_1 | Q_2) := \left\{ \begin{array}{ll} \int dQ_2 \left(
\frac{dQ_1}{dQ_2} \log \frac{dQ_1}{dQ_2} \right) & \mbox{if } Q_1
\ll Q_2 \mbox{ and } \frac{dQ_1}{dQ_2} \log \frac{dQ_1}{dQ_2} \in
L^1(Q_2) \\ +\infty & \mbox{otherwise}
\end{array} \right.
\]
the relative entropy of $Q_1$ with respect to $Q_2$. \bt{t1} The
sequence $P_N \circ \rho_N^{-1}$ of elements of $\M_1$ satisfies a
{\em Large Deviation Principle} (LDP) with good rate function
\[
I(Q) := H(Q | W \otimes \mu) - F(Q) .
\]
\et The proof of Theorem \ref{t1}, as well as of the other results stated in this section,  is postponed to the appendix.\\
We recall that the above statement means that, for each Borel
subset $A$ of $\M_1$,
\[
- \inf_{Q \in {\buildrel _{\circ} \over {\mathrm{A}}}} I(Q) \leq \liminf_{N \ra +\infty} \frac{1}{N}
\log P_N \circ \rho_N^{-1} (A) \leq \limsup_{N \ra +\infty}
\frac{1}{N} \log P_N \circ \rho_N^{-1} (A) \leq - \inf_{Q \in
\overline{A}} I(Q),
\]
where ${\buildrel _{\circ} \over {\mathrm{A}}}$ and $\overline{A}$ denote the interior and the
closure of $A$ respectively; moreover the function $I(\cdot)$ is
nonnegative, lower-semicontinuous, and the level sets $\{Q:I(Q) \leq
l\}$ are compact, for each $l>0$. \bt{t2} The equation $I(Q) = 0$
has a unique solution $Q_*$, that can be identified as follows.
Consider the nonlinear integro-differential equation \be{MKV}
\left\{ \begin{array}{rcl} \frac{\partial}{\partial t} q_t(\l) & = &
e^{-\g} \exp\left[ \b \int \mu(d\l') \a' q_t(\l') \right]
(1-q_t(\l)) \\ q_0(\l) & \equiv & 0
\end{array} \right.
\ee for a real-valued $q_t(\l)$, $t \geq 0$, $\l = (\a,\b,\g) \in
\R^3$. This equation has a unique solution $0 \leq q_t(\l) \leq 1$.
For every $\l$ fixed, consider the Markov chain on $\{0,1\}$ with
time-dependent infinitesimal generator \be{Gq} L_t^{\l} f(s) :=
c_t^{\l}(s) [f(1-s) - f(s)], \ee where
\[
c_t^{\l}(s) := (1-s) e^{-\g} \exp\left[ \b \int \mu(d\l') \a'
q_t(\l') \right]
\]
and starting from $s=0$ (note that this process jumps only once,
from $0$ to $1$, and it is then trapped in $1$). Let $Q_*^{\l}$ be
the law of this process on $\D[0,T]$. Then
\[
Q_* = Q_*^{\l} \otimes \mu.
\]
Moreover
\[
q_t(\l) = Q_*^{\l} (\s(t) = 1).
\]
\et Theorems \ref{t1} and \ref{t2} have a simple consequence. Let
$U$ be an open neighborhood of $Q^*$ in $\M_1$. By Theorem \ref{t2},
lower semicontinuity of $I(\cdot)$ and compactness of its level
sets, a standard argument shows that $k(U) := \inf_{Q \not\in
\overline{U}} I(Q) > 0$. By the upper bound in Theorem \ref{t1}
there exists $C>0$ such that
\[
P_N (\rho_N \in U) \leq C e^{-N k(U)},
\]
thus converges to zero with exponential rate. We summarize this fact
in the following {\em law of large numbers}. \bc{c1} Let
$d(\cdot,\cdot)$ be any metric that induces the weak topology on
$\M_1$. Then for every $\e>0$, the probability
\[
P_N (d(\rho_N,Q_*) \geq \e)
\]
converges to zero with exponential rate in $N$. \ec Next result is
about the fluctuations of $\rho_N$ about $Q_*$, which has the form
of a Central Limit Theorem. In most cases, these dynamical
fluctuation theorems are proved by method of weak convergence of
processes. A typical tool in this context is Theorem 1.6.1 in \cite
{EK}; it has been widely applied to models close in spirit to this
work, see for instance \cite{CE} and \cite{DRST}. The effectiveness
of those methods for heterogeneous models is, however, unclear. The
main point is that via Theorem 1.6.1 in \cite {EK}
 one obtains the dynamics of the fluctuation process, which is infinite dimensional;
 to get a {\em computable} expression for the asymptotic variance of a given function
 of the trajectory may be not feasible.

\noindent We follow here a different approach, which allows to prove
a Central Limit Theorem directly in the space $\M$. This is inspired
by the seminal work of E. Bolthausen \cite{B}, and it has been
carried out in a context similar to ours in \cite{BB} and
\cite{DDD}. The main difference here is that the underlying
stochastic dynamics $P_N^{\l}$ are not reversible; the related
difficulties have forced us to introduce a further assumption, that
we call {\em reciprocity condition}: \bi
\item[(R)] There exists a deterministic $b>0$ such that for all $i
\geq 1$ the identity $\b_i = b \a_i$ holds almost surely. \ei In
economical terms, this means that the sensibility of a firm to
variation of the aggregate variable $m_N$ is proportional to the
impact the default of that firm has in the network. In different
words, the interaction between firms is {\em symmetric}: if the
$i$-th firm strongly interact with the network (i.e. $\a_i$ is
large) then it has a large influence on the network but,
symmetrically, it is also strongly influenced by the state of the
network. This assumption may be reasonable in many situations.

\noindent From now on, whenever condition (R) is assumed, $(\l_i)$
will denote the pair $(\a_i,\g_i)$; accordingly, $\M_1$ and $\M$
will be spaces of measures on $\D[0,T] \times Supp(\mu)$, where
$Supp(\mu)\subset \mathbb R^+ \times \mathbb R$.

\noindent In order to state our Central Limit Theorem, we need to
introduce some notations. Let $\M_0$ be the subset of $\M$ comprised
by the signed measures with zero total mass. Let $\nu_*$ be the law,
induced by $Q_*$, of the $\M_0$-valued random variable
$\d_{(\s[0,T],\l)} - Q_*$. We then denote by ${\cal{C}}_b$ the space
of bounded, continuous, real valued functions on $\D[0,T] \times
\R^2$. For $\phi \in {\cal{C}}_b$, define $\hat{\phi} \in \M_0$ by
\be{Hp} \hat{\phi}(A) := \int  \nu_*(dR) \left( R(A) \int \phi dR
\right), \ee for $A \subseteq \D[0,T] \times \R^2$ measurable.
\bt{t3} Let $\phi_1,\phi_2,\ldots, \phi_n \in {\cal{C}}_b$. Then the
$P_N$-law of the random vector
\[
\sqrt{N} \left( \int \phi_i d\rho_N - \int \phi_i dQ_*
\right)_{i=1}^n
\]
converges weakly as $N \ra +\infty$ to an $n$-dimensional Gaussian
probability measure with zero mean and covariance matrix $C =
(C_{i,j})_{i,j =1}^n$ given by
\[
C_{i,j} := \int (\phi_i - \phi_i^*)(\phi_j - \phi_j^*)dQ_* - D^2
F(Q_*) [ \hat{\phi}_i,  \hat{\phi}_j],
\]
where $\phi_i^* := \int \phi_i dQ_*$ and $D^2 F(Q_*) [
\hat{\phi}_i, \hat{\phi}_j]$ is the second Fr\'echet directional
derivative of $F$ at $Q_*$ in the directions $\hat{\phi}_i,
\hat{\phi}_j$:
\begin{eqnarray*}
DF(Q_*) [ \hat{\phi}_i] & = & \lim_{h \ra 0} \frac{F(Q_* + h \hat{\phi}_i) - F(Q_*)}{h} \\
D^2 F(Q_*) [ \hat{\phi}_i,  \hat{\phi}_j] & = &  \lim_{h \ra 0}
\frac{DF(Q_* + h \hat{\phi}_j) [ \hat{\phi}_i] - DF(Q_*) [
\hat{\phi}_i] }{h}
\end{eqnarray*}
(all these limits will be shown to exist).\\
Moreover the diagonal terms of the covariance $C_{i,j}$ can be
written as \be{CovDiag} C_{i,i}
 =
E^{Q_*}\left[\left((\phi_i-\phi^*_i)-\b\int_0^T(1-\s(s))
Cov_{Q_*}(\a\s(s),\phi_i) dM(s)\right)^2\right], \ee where
\be{Mart}M(t):=\ind_{\{\tau\leq t\}}-\int_0^{t}(1-\s(s)) e^{-\g+
\b\int \a'\eta(s) Q_*(d\eta[0,T],d\l') }ds\ee is the compensated
$(Q_*,\cal{F})$-martingale associated with the jump process of
$\s[0,T]$.\et

\section{Applications to the portfolio analysis}
\subsection{Computation of large portfolio losses}
We are now going to state a definition of \emph{portfolio losses}.
When speaking of portfolio losses, we mean the losses that a
financial institution
 may suffer in a credit portfolio due to the default events.
 Many specifications may be chosen to this aim. Some general rules are now
 stated. A rather general modeling framework is to consider the total loss
that a bank may suffer due to a risky portfolio at time $t$ as a
random variable defined by $L^{N}(t)=\sum_{i}L_{i}(t)$, where
$L_{i}(t)$ is the loss, called {\em marginal loss}, due to the
obligor $i$. Different specifications for the marginal losses
$L_{i}(t)$ can be chosen accounting for heterogeneity, time
dependence, interaction, macroeconomic factors and so on. A punctual
treatment of this general modeling framework can be found in the
book by Embrechts, Frey and McNeil \cite {FMN}. For a comparison
with the most widely used industry examples of credit risk models
see \cite{FM2} or  \cite{CGM}. The same modeling insights are also
developed in the most recent literature on risk management and large
portfolio losses analysis, see \cite{GW}, \cite{FB} and \cite{DDDD}
for different specifications.

Here we assume that \be{Li}L_i(t):=\varphi(\s_i[0,t],\l_i,t),\ee
where $\varphi(\, \cdot \, , \, \cdot \, , t)$ is bounded and
continuous in ${\cal{D}}[0,t] \times \mbox{supp}(\mu)$. In other
words the marginal loss depends explicitly on the realization of
$\l_i$ and on the
\emph{history} of $\s_i$.\\
As a particular case of our general framework we obtain the most
standard set up commonly used in the literature of credit risk:
consider $\varphi(\s_i[0,t],\l_i,t):=   e(\l_i,t) \s_i(t)$ where
$e: \R^3 \to \R^+$  is a continuous function of $\l_i$, and measures the \emph{exposure} in case of
default. Thus
\[ L^{(N)}(t)=\sum_{i=1}^N e(\l_i,t) \s_i(t).\]
We shall often speak of  \emph{asymptotic loss} or \emph{asymptotic
portfolio}. In this case we are referring at the $N\to\infty$ case
of infinite obligors. The large portfolio is intended to be a
\emph{large} but finite approximation of this asymptotic regime.

As a consequence of the central limit theorem for the empirical
measure, we obtain the following description for $L^N(t)$, the
aggregate losses computed at time $t\in[0,T]$:

\bc{loss1} Assume the reciprocity condition (R) is satisfied, and consider $L_i(t)$ as given in (\ref{Li}). In what follows, for $y[0,t] \in {\cal{D}}[0,t]$ and $\l \in \mathbb R^+ \times \mathbb R$, we write $L(t)$ for $\varphi(y[0,t],\l,t)$, and $l(t)=E^{Q_*}[L(t)]$. As $N\to\infty$
the sequence
$$\sqrt{N}\left[\frac{\sum_{i=1}^NL_i(t)}{N}-l(t)\right]$$
 converges weakly to a centered
Gaussian random variable with variance $V(t)$ where \be{VT0}
V(t)=E^{Q_*}\left[ \left(  L(t)-l(t) - \b \int_0^{t} (1-\s(s))
Cov_{Q_*}(\a\s(s),L(t))\ dM(s)  \right)^2\right]\ee and where
$M(t)$ has been defined in (\ref{Mart}).

 \ec
\bpr We apply Theorem \ref{t3}
with
$n=1$ and $\phi_1 = \phi=\varphi(\s[0,t],\l,t)=L(t)$. In this case
$\int \phi d\rho_N=\frac{L^N(t)}{N}$ and $\int \phi dQ_*
=E^{Q_*}[L(t)]=l(t)$. Notice that we can consider, without loss of
generality, $t$ as the final horizon of the time period, i.e. $t =
T$. \epr

 \br{newremark}
 By applying Theorem \ref{t3}  with $(\phi_j)_{j=1}^n$, where $\phi_j := \varphi(y[0,t_j],\l,t_j)$,
 one could show that the {\em finite dimensional distributions} of the process
 \[
 \left(\sqrt{N}\left[\frac{\sum_{i=1}^NL_i(t)}{N}-l(t)\right] \right)_{t \geq 0}
 \]
 converge to the ones of a Gaussian process, whose covariance can in principle be computed. Our Theorem \ref{t3} does not allow, however, to prove weak convergence of the process.
  \er

The asymptotic expected  value $l(t)$ corresponds to the fraction of
loss at time $t$ in a benchmark portfolio of infinite firms.
Concerning equation (\ref{VT0}), notice that in the case of no
interaction, (i.e., $\beta=0$) we have
$$V(t)=E^{Q_*}\left[ (L(t)-l(t))^2\right]=Var_{Q_*}(L(t)).$$
In the case of $\beta>0$ there is a suppletive noise given by the
interaction.
 It depends on the  past history of the process, hence a sort of ``\emph{memory}"
 of the variance $V(t)$.\\

The variance given in (\ref{VT0}) involves the integral with respect to a martingale.
A simpler form for the variance, more suitable for numerical computations, can be found in the following special but significant
case.
\bp{P1} Suppose that
$L_i(T)=e(\l_i,T)\s_i(T)$. Then as
$N\to\infty$ we have that
$$\sqrt{N}\left[\frac{L^N(T)}{N}-l(T)\right]$$
where $l(T)=E^{Q_*}[e(\l,T)\s(T)]$, converges to a centered Gaussian
random variable with variance \be{VT}V(T)=Var_{Q_*}(L(T)) +
\int_0^{T} \left( E^{Q_*} [\a  \s(s)(e(\l,T)-l(T)) ]\right)^2 \cdot
E^{Q_*} [\b^2 (1-\s(s))e^{-\g+\b m_{Q_*}(s)}] \ ds,\ee where
$m_{Q_*}(t):=\int\a'\eta(t)Q_*(d\eta[0,T],d\l')$.\ep

\bpr We only need to show that the variance can be written in the
form given in (\ref{VT}). From (\ref{VT0}) it is easy to see that
$V(T)$ can be written as \be{VT1}V(T)=E^{Q_*}\left[
\left(L(T)-l(T)-  \int_0^{T }\b (1-\s(s)) E^{Q_*} [\a \s(s)
(e(\l,T)\s(T)-l(T)) ] \ dM(s)\right)^2 \right].\ee We now look at
the expectation in the integral $$ E^{Q_*} [\a \s(s)
(e(\l,T)\s(T)-l(T)) ] =
 E^{Q_*}[\a\s(s)(e(\l,T)-l(T))].
$$
where we have used the fact that $E^P[\s(s)\s(T)]=E^P[\s(s)]$ for
$s\leq T$ and for all $P\in\M_1$.  Substituting in (\ref{VT1}) we
have
\begin{multline}\label{VT3} V(T)=E^{Q_*}\left[ \left(L(T)-l(T) -
\int_0^{T}\b (1-\s(s))  E^{Q_*}[\a\s(s)(e(\l,T)-l(T))]\ dM(s)
\right)^2 \right]\\=
   E^{Q_*}\left[ (L(T)-l(T))^2 \right] -\\- E^{Q_*}\left[  2
(L(T)-l(T))   \int_0^{T } \b
(1-\s(s))E^{Q_*}[\a\s(s)(e(\l,T)-l(T))]\ dM(s) \right]  +\\+
E^{Q_*}\left[ \left(   \int_0^{T}\b(1-\s(s))
E^{Q_*}[\a\s(s)(e(\l,T)-l(T))]\ dM(s) \right)^2\right]
\end{multline}
The first expectation is the variance of $L(T)$ computed under
$Q_*$. We show now that the second expectation is null. Indeed it
is equal to
\begin{multline*} 2  E^{Q_*}\left[    L(T)    \int_0^{T }  \b (1-\s(s))
E^{Q_*}[\a\s(s)(e(\l,T)-l(T))]\ dM(s)  \right]\\-2   l(T)
E^{Q_*}\left[
 \int_0^{T }  \b (1-\s(s))  E^{Q_*}[\a\s(s)(e(\l,T)-l(T))]\ dM(s)  \right]
 \end{multline*}
 The second term is zero since $M $ is a $(Q_*;\cal{F})-$martingale
 and the argument of the integral is
 $\mathcal{F}_s$ measurable.  Concerning the first one we
see that \begin{multline*}E^{Q_*}\left[    L(T)    \int_0^{T}
(1-\s(s))(\cdot)\ dM(s) \right] = E^{Q_*}\left[ e(\l,T)
\ind_{\{\tau\leq T\}} \int_0^{T\wedge \tau} (\cdot)\ dM(s) \right]\\
= E^{Q_*}\left[ e(\l,T)
    \int_0^{\tau} (\cdot)\ dM(s)  \right]=0.\end{multline*}
Where the last equality is due to the fact that $e(\l,T)$ is
$\mathcal F_0$ measurable. Concerning the last term in (\ref{VT3})
we   now show that \be{isom} E^{Q_*}\left[ \left(
\int_0^{T}\b(1-\s(s)) E^{Q_*}[\a\s(s)(e(\l,T)-l(T))]\ dM(s)
\right)^2\right]=\ee$$ =  E^{Q_*} \left[ \int_0^{T}\left[ \b
(1-\s(s)) E^{Q_*}[\a\s(s)(e(\l,T)-l(T))]\ \right]^2  (1-\s(s))
e^{-\g+\b m_{Q_*}(s)} d s\right].
$$ Indeed, being  $M(t)=\ind_{\{\tau\leq t\}}-\int_0^t (1-\s(s))
e^{-\g+\b m_{Q_*}(s)} ds$, it can be shown that the quadratic
variation $\langle M\rangle$ of $M$, is $\langle
M\rangle_t=\int_0^t (1-\s(s)) e^{-\g+\b m_{Q_*}(s)} ds$, for all
$t\in[0,T]$. Thus for each progressively measurable process
$X=(X(t))_{t\in[0,T]}\in L^2_M([0,T])$, where
$L^2_M([0,T]):=\left\{X : E^{Q_*}\int_0^T|X(s)|^2 d\langle
M\rangle_s <\infty\right\}$,
$$\left\| X \right\|_{L^2_M([0,T])}=E^{Q_*} \int_0^T |\ X(s)|^2 d\langle M\rangle_s=
E^{Q_*}  \int_0^{T } |X(s)|^2 (1-\s(s))  e^{-\g+\b m _{Q_*}(s)} ds
.$$ As a consequence, applying this last result with
$X=(X(t))_{t\in[0,T]}$ defined as
$$X(t)=  \b(1-\s(t)) E^{Q_*}[\a\s(t)(e(\l,T)-l(T))]dt,$$ by
the isometry   between $L^2(\Omega,\mathcal F_T,Q_*)$ and
$L^2_M([0,T])$, we have
$$\left\|\int_0^T X(s) dM(s) \right\|_{L^2(\Omega,\mathcal F_T,Q_*)}=\left\| X \right\|_{L^2_M([0,T])}$$
which is exactly (\ref{isom}). Finally
\begin{multline*}E^{Q_*} \int_0^{T}\left[ \b (1-\s(s))
E^{Q_*}[\a\s(s)(e(\l,T)-l(T))]\ \right]^2 (1-\s(s)) e^{-\g+\b
m_{Q_*}(s)} d s
\\= \int_0^{T}
\left( E^{Q_*}[\a\s(s)(e(\l,T)-l(T))] \right)^2 \cdot
E^{Q_*}[\b^2(1-\s(s)) e^{-\g+\b m_{Q_*}(s)} ] \ d s, \end{multline*}
and this proves (\ref{VT}).

 \epr

In the next section we show an application of this law of large
number (Theorem~\ref{t2}) and central limit theorem (Proposition
\ref{P1}). Indeed, we show the evolution of the  default probability
of certain groups of obligors and infer the corresponding
probability of suffering large losses in a credit portfolio.
\subsection{An example with simulation results}\label{ex1}
\begin{figure}[t]
\begin{center}
\includegraphics[height=9cm]{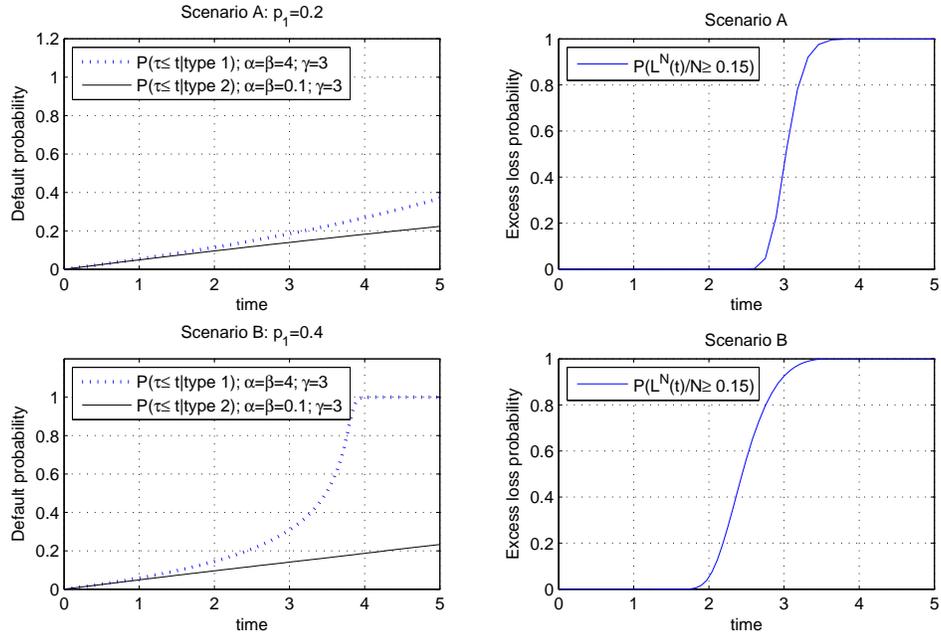}
\caption{{\small{Conditional probabilities of default and excess
loss probabilities in a portfolio of $N=125$ obligors where
$L^N(t)$ is as defined  in Section \ref{ex1}. Scenario A
represents a portfolio where only the $20\%$ of obligors are of
type 1 whereas in Scenario B $40\%$ of the obligors are of type 1.
In both Scenario A and B we have $\lambda_1=[4,4,3]$,
$\lambda_2=[0.1,0.1,3]$, where $\lambda_i$ describes the
idiosyncratic characteristics of obligors of type $i=1,2$.}}}
\label{plot1_2}
\end{center}
\end{figure}

\begin{figure}[h]\begin{center}
\includegraphics[height=9cm]{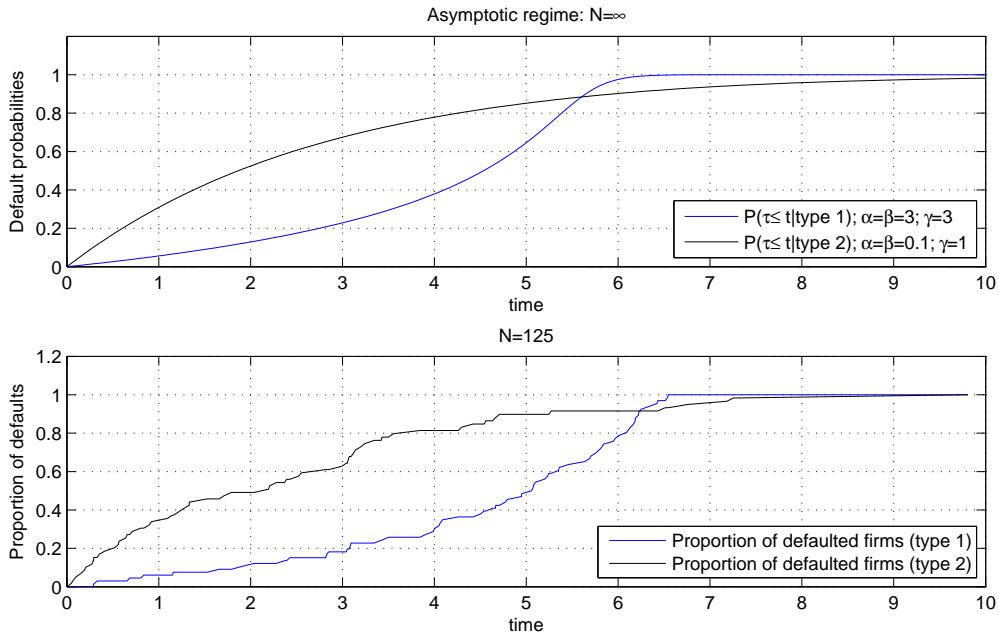}
\caption{{\small{Comparison between the dynamics of the default
probabilities of two groups of firms computed under two different
models. In the upper part we see the plot under the asymptotic
model with infinite firms (under $Q_*$). In the lowest one we have
implemented a simulation of the Markov process directly (here
$N=125$). The parameters are $\lambda_1=[3,3,3]$,
$\lambda_2=[0.1,0.1,1]$, where $\lambda_i$ describes the
idiosyncratic characteristics of obligors of type $i=1,2$. The
distribution of $\l$ is in this case
$\mu=\frac{1}{2}\d_{\l_1}+\frac{1}{2}\d_{\l_2}$.}}} \label{plot9}
\end{center}
\end{figure}

 Consider the simplified case in which $L_i(T)=\s_i(T)$
and zero otherwise: the exposure at default is equal to one for each
obligor. Moreover assume that
$\mu=p_1\delta_{\l_1}+(1-p_1)\delta_{\l_2}$. This means that the law
of the random environment $\l=(\alpha,\beta,\gamma)$ puts mass on
two possible outcomes, that is $\l\in\{\l_1;\l_2\}$. In this case
the index $i=1,2$ identifies a group of obligors with the same
marginal characteristics. In other words, we split the portfolio in
two types of obligors with different specifications. More complex
specifications could also be chosen since, as already said in the
introduction, we are not forced to split the sample in groups with
homogeneous characteristics. For sake of simplicity we start we an
illustrative example where interesting
features on the dynamic of the state variables can be captured.  \\
Recall that $\alpha_i$ specifies the relative weight of firm $i$ in
building the aggregate variable $m_N=\frac 1N \sum_i \a_i \s_i$ (see
equation (\ref{generator})). $\beta_i$ is the parameter that
measures how sensitive  obligor $i$ is with respect to the aggregate
variable $m_N$, put differently it is a measure of the contagion
effect. $\gamma_i$ is the idiosyncratic term in the marginal default
probability.\\
As an example, we take a portfolio of $N=125$ obligors. This is a
typical size for CDO's portfolios. We suppose that the portfolio
consists of obligors of two types, in particular $\l_1=(4,4,3)$,
$\l_2=(0.1,0.1,3)$. Notice that $\a_i=\b_i$ hence the reciprocity
condition (R) applies and we
are allowed to rely on Theorem \ref{t3} and Corollary \ref{loss1}.\\
Obligors of type 1 are more sensitive to the aggregate variable,
that is, their marginal default probability depends strongly on the
default indicator of the other firms. Obligors of type 2 are less
influenced by the aggregate variable $m_N$. The idiosyncratic term
$\gamma$ is the same for each obligor.\\
With this choice of the parameters, we want to stress the fact that
even though the marginal default probabilities of the two types are
rather similar for the very short horizon (where the impact of
$\gamma$ is higher), the contagion effect becomes preeminent as time
goes by, at least
under certain specifications in the construction of the portfolio.\\
To illustrate this situation, in Figure \ref{plot1_2} (on the left),
we show the dynamics of the marginal default probability of the two
groups in two different scenarios. Scenario A mimics a portfolio
where only $20\%$ of the obligors are in group 1, this means that
the proportion of firms exposed to \emph{contagion risk} is lower.
In scenario B the proportion is increased to $40\%$. Notice that in
the second scenario the probability of default of the firms in the
first group increases dramatically after the second year. Firms of
type 2 are less influenced.\\
Relying on Proposition \ref{P1},  we compute  the corresponding
excess probabilities, that is, the probability of suffering a loss
bigger than $x$ as a function of time in the two scenarios. In
Figure \ref{plot1_2} (on the right) we see this probabilities for
$x=0.15$. Notice that in this simple example, $L^N(t)$ counts the
number of defaults up to time $t$, so that $P(L^N(T)/N\geq 0.15)$
represents the probability of having at least $15\%$ of defaults in
the whole portfolio. Looking at the graphs, it is easy to see that
in the first scenario the probability of having such a loss is
smaller. At time $t=2.5$ the $P(\frac{L^N(2.5)}{N} \geq x)\sim 0$
whereas in the
second scenario $P(\frac{L^N(2.5)}{N} \geq x)\sim 0.55$.\\
This simple example suggests that the contagion effect can be very
significant when looking at the probability of suffering a certain
loss in a large portfolio. This is crucial for risk measurements
purposes and for pricing trances of CDO's.\\
It may be argued that an approximation via an infinite portfolio may
not reproduce the real situation.  In Figure \ref{plot9} we show a
comparison between the evolution in time of the default
probabilities in the two groups of obligors computed under $Q_*$ (in
the upper part) and simulated via the \emph{real} Markov process
with $N=125$ (in the lower part). Here the parameters are
$\l_1=(3,3,3)$, $\l_2=(0.1,0.1,1)$. These graphs show that the
asymptotic equation is a good approximation for the Markov process even with $N=125$.\\
\begin{figure}[h]
\begin{center}
\includegraphics[height=9cm]{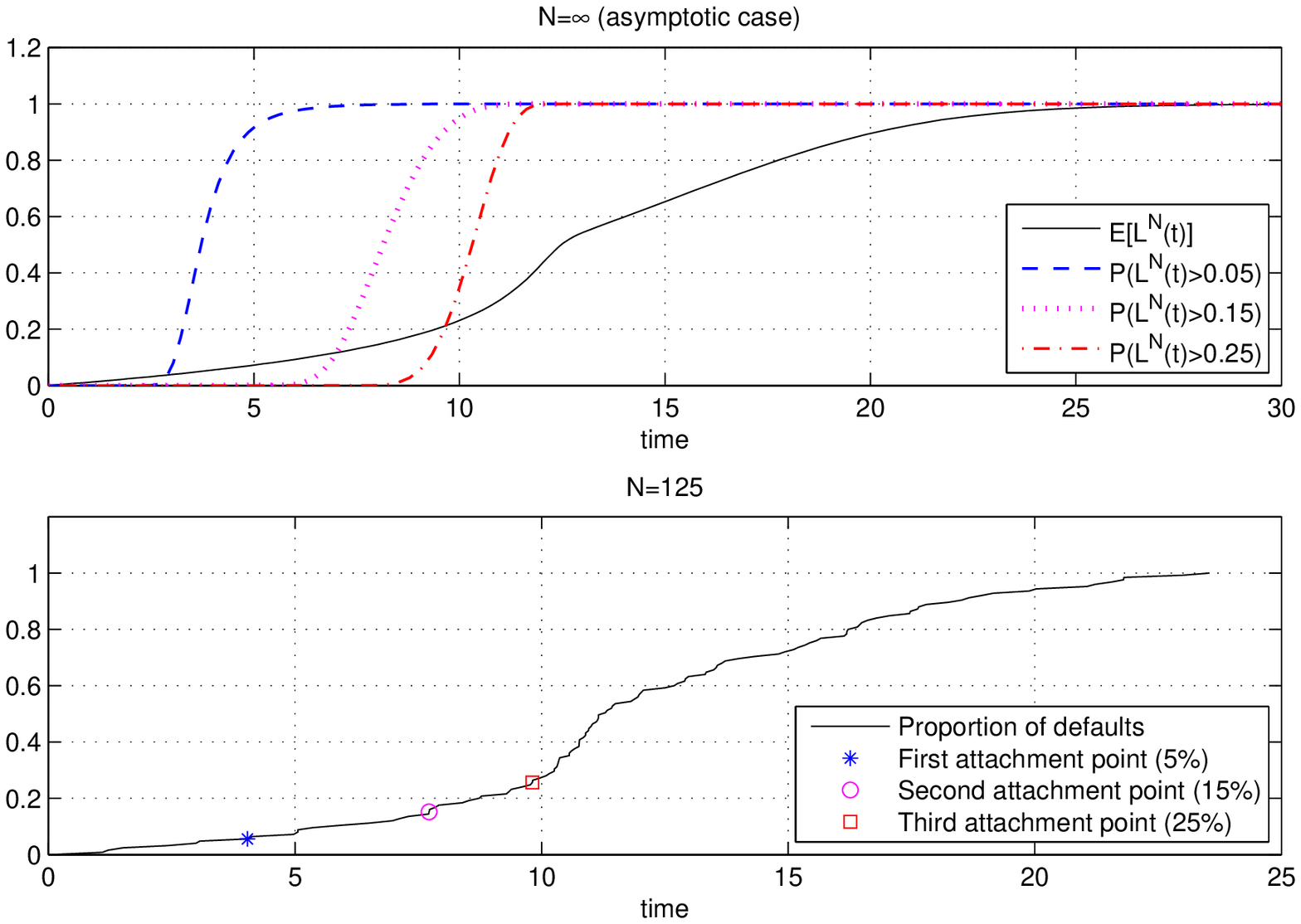}
\caption{{\small{Evolution in time of the aggregate losses (in
black) computed under two different models: under the asymptotic
model (upper) and  the Markov process (lower). The parameters are
$\lambda_1=[2,6,4]$, $\lambda_2=[1,3,5]$, where $\lambda_i$
describes the idiosyncratic characteristics of obligors of type
$i=1,2$. The distribution of $\l$ is in this case
$\mu=0.4\d_{\l_1}+0.6\d_{\l_2}$.}}} \label{plot10}
\end{center}
\end{figure}
Finally we show in Figure \ref{plot10} the comparison between the
evolution of the aggregate loss (in black) evaluated as before in
two different ways. In the upper part we see the aggregate loss
computed relying on Proposition \ref{P1}. Below we see the plot of a
trajectory of the Markov Process (with $N=125$). In the same graphs
we have also plotted the dynamics of the probability of suffering a
loss over certain thresholds $c_k$. In other words we plot $t\mapsto
P(L^N(t)\geq c_k)$ for $k=1,2,3$ (in this case $c_1=5\%, c_2=15\%,
c_3=25\%$). Those probabilities are building blocks for computing
the price of trances of CDO's contracts. For a description of such a
credit derivative see for instance \cite{FMN}.

\section{Conclusions}
We have proposed a  model for \emph{credit contagion} where
\emph{heterogeneity} and \emph{direct contagion} among the firms are
taken into account. We have then quantified the impact of contagion
on the losses suffered by a financial institution holding a large
portfolio with positions issued by the firms.

Compared to the existing literature on credit contagion, we have
proposed a \emph{dynamic} model where it is possible to describe the
evolution of the indicators of financial distress. In this way we
are able to compute the distribution of the losses in a large
portfolio for any time horizon $T$, via a suitable version of the
central limit theorem.


The peculiarity of our model is the fact that the homogeneity
assumption is broken by introducing a \emph{random environment} that
makes it possible to take into account  the idiosyncratic
characteristics of the firms. One drawback of the intensity based
models commonly proposed in the literature  is the difficulty in
managing large heterogeneous portfolios because of the presence of
many obligors with different specifications. In this case it is
common practice to assume homogeneity assumptions in order to reduce
the complexity of the problem. A typical approach is to divide the
portfolio into groups where the obligors may be considered
exchangeable. We have shown  that our model goes behind the
identification of groups of firms that can be considered basically
exchangeable. Despite this heterogeneity assumption our model has
the advantage of being totally tractable: it is possible to compute
in closed form the mean and the variance of a central limit type
approximation for the losses due to a large portfolios in a dynamic
fashion.

As an example of the general theory we have computed the default
probabilities and different risk measures in a simple situation with
only two groups of obligors. Moreover we have compared the numerical
results obtained relying on the asymptotic model and on the central
limit theorem (Corollary \ref{loss1}) with the results obtained in a
simulation of the underlying Markov process with finite $N=125$.
These results show the goodness of the approximation and are
encouraging a more involved analysis. This issue is left to future
research: it is in fact out of the scope of this work to pursue a
punctual calibration of the model to real data.

\appendix
\section{Proofs of main results}
\subsection{Proof of Theorem \ref{t1}}

We need to prove some technical lemmas.

\bl{varad2} Let $\mathcal X$ be a Polish space. Let $(P_N)_N$
satisfy the LDP with rate $N$ and good rate function $H$. Let
$F:\mathcal X\to \mathbb R$ be measurable, bounded from above and
continuous on the set $\mathcal X_H:=\{x:H(x)<\infty\}$. Then the
sequence of probability measures $(P^F_N)_N$ defined by \be{3.8}
\frac{dP^F_N}{dP_N}(\cdot)=\frac{ \exp{(N F(\cdot))}}{\int_{\mathcal
X}\exp{(N F(y))}P_N(dy)} \ee satisfies the LDP with the good rate
function \be{I2} I(x)=H(x)-F(x)-\inf_{y\in\mathcal X}[H(y)-F(y)].\ee
In particular
\begin{equation}\label{lub4}
\lim_{\,N\rightarrow+\infty}\frac{1}{N}\,\log\left[\int_{\mathcal X}
\exp\left(N F(y)\right)P_N(dy)\right]=-\inf_{\,y\, \in\, \mathcal
X}\left[H(y)-F(y)\right].
\end{equation}
\el For a proof see \cite{T}. This is a relaxed version of the usual
Varadhan's Lemma for tilted large deviations principles (see
\cite{VA}). The statement is relaxed in the sense that we assume
that a suitable function $F:\mathcal X\to \mathbb R$,
instead of being continuous on all its domain,
is continuous only on a subset $\mathcal X_H\varsubsetneq \mathcal X$ in the following sense:\\
\emph{For any sequence $(x_n)_n\in \mathcal X$ such that $x_n\to x$, where $x\in\mathcal X_H$ we have $F(x_n)\to F(x)$. }\\
We point out that this is a stronger assumption than assuming
continuity of the restriction of $F$ on the subset $\mathcal X_H$.

\bl{l1} For given $\ul\in(\R^+\times\R^+\times\R)^N$,
\be{gulp2}\frac{dP^{\ul}_N}{dW^{\otimes
N}}(\us[0,T])=\exp\{NF(\rho_N(\us[0,T],\ul))\}.\ee where $F(Q)$ has
been defined in  (\ref{F}).\el \bpr It basically follows from  the
Girsanov formula for point processes (See \cite{Br}).
$$\frac{dP^{\ul}_N}{dW^{\otimes N}}=\exp\left\{ \sum_{i=1}^N
\int_0^T\left[  (1-\s_i(t))- (1-\s_i(t))e^{-\g_i+\b_i\int \a'\eta(t)
\r_N(d\eta[0,T],d\l')} \right]dt+ \right.$$
$$\hskip +1cm + \left. \sum_{i=1}^N \s_i(T){\left[-\g_i+\left(\b_i\int \a'\eta(t^-)
\r_N(d\eta[0,T],d\l')\right)\Big|_{t=\tau_i}\right]} \right\} ;$$
where  $\tau_i$ has been defined in (\ref{deft}).\\ The term in
the  $\{\ \}$-brackets is exactly $F(Q)_{|Q=\rho_N}$ multiplied by
$N$. \epr

We now define for each $Q\in\M$ and $t\in[0,T]$
\be{Def_m}m_Q(t):=\int \alpha \eta(t) Q(d\eta[0,T],d\l).\ee We
shall often use this notation in the rest of this appendix.

\bl{L313} F(Q) is bounded on   $ \M_1$ and continuous on the
subset $\M_A:=\{Q\in\M_1:Q\ll W\otimes \mu \}$ \el \bpr We rewrite
$F(Q)$ as given in (\ref{F}) using  the notation introduced in
(\ref{Def_m}).

 \begin{multline}\label{FFF}
 F(Q)  = \int Q(d\s[0.T],d\l) \left\{ \int_0^T (1-\s(t)) \left( 1-
e^{-\g} e^{\b m_Q(t)} \right) dt \right.
\\ + \left. \s(T) \left[ - \g +  \b m_Q(t^-) \Big|_{t=\tau(\s[0,T])} \right]
\right\},
\end{multline}
The argument in the  $\{\ \}$-brackets   is bounded, thus we are
allowed to interchange the expectation with respect to $Q$ and the
time
 integral.

\begin{multline*}
F(Q)=\int_0^T E^Q \left[1- \s(t)\right]\ dt-\int_0^T
 E^Q\left[ e^{-\g}e^{\b m_Q(t)}\ (1-\s(t))
\right] \ dt -\\-  E^Q\left[\g \s(T) \right]+ E^Q\left[\b \s(T)
m_Q(t^-)\Big|_{t=\tau(\s[0,T])} \right].\end{multline*}

\noindent We show now boundedness and continuity of $F$. The
boundedness is easily proved since $\s\in\{0;1\}$, and the
distribution of $\l$ under $Q\in\M_1^b$ has bounded support.\\
In order to prove the continuity on $\mathcal M_A$, we consider a
sequence of probabilities $(Q_n)_{n\geq 0}\in \mathcal M_1^b$
converging weakly to $Q\in\mathcal M_A$. We split the proof in
different steps.\\
We    show first that \be{canc}\lim_n
E^{Q_n}[f(\l)\s(t)]=E^{Q}[f(\l)\s(t)]\ee for all $t\in[0,T]$ and
for any  continuous function $f:\mathbb R^+ \times \mathbb R^+ \times \mathbb R \to \mathbb R$ bounded on
the support of $\mu$. This statement is not trivial, since the
projection $\s{[0,T]}\to \s(t)$ is not continuous in $\D[0,T]$.
However, define for any $\e>0$ the functions
$$g^{-}_t(\e;f):=\frac{1}{\e}\int_{t-\e}^{t}f(\l)\s(s)ds\ ,\qquad
g^{+}_t(\e;f):=\frac{1}{\e}\int_{t}^{t+\e}f(\l)\s(s)ds;$$
where we suppose that the trajectory $\s[0,T]$ can be extended to the larger interval $[0-\e,\ T+\e]$ by continuity.\\
These functions are continuous in $\D[0,T]$, bounded by $\|f
\|_\infty$ for any $\e$ and such that $g^{-}_t(\e;f)\leq
f(\l)\s(t)\leq g^{+}_t(\e;f)$ a.s. for any $t$. Thus, by the
Lebesgue convergence theorem,
$$\limsup_n E^{Q_n}[f(\l)\s(t)]\leq \lim_n E^{Q_n}[g^+(\e;f)]= E^Q[g^+(\e;f)], \quad \forall \e>0.$$
Letting $\e\to 0$ and noticing that  $\lim _{\e\to 0}
g^{+}_t(\e;f)=f(\l)\s(t)$ we get
$$\limsup_n E^{Q_n}[f(\l)\s(t)]\leq E^Q[f(\l)\s(t)].$$
The same argument holds for $g^{-}_t(\e;f)$; here  $\lim _{\e\to 0}
g^{-}_t(\e;f)=f(\l)\s(t^-)$. Thus
$$E^Q[f(\l)\s(t^-)]\leq \liminf_n  E^{Q_n}[f(\l)\s(t)] \leq \limsup_n E^{Q_n}[f(\l)\s(t)]\leq E^Q[f(\l)\s(t)].$$
Notice that $f(\l)\s(t)$ and $f(\l)\s(t^-)$ may differ only on the
event $\{\s(t^-)\ne\s(t)  \}$. But this event has measure zero for
any $Q\in\mathcal M_A$, since $(W\otimes \eta)(\{\s(t^-)\ne\s(t)
\})=0$. This implies that the corresponding expected values must
coincide; as a consequence
$E^Q[f(\l)\s(t)]-E^Q[f(\l)\s(t^-)]=0$.
We have thus proved that
 \be{mtom2}\lim_n E^{Q_n}[f(\l)\s(t)]=E^{Q}[f(\l)\s(t)] \ \ \mbox{for all}\
 t.\ee
  Notice that in saying that
$(W\otimes \eta)(\{\s(t^-)\ne\s(t)  \})=0$ we have used the fact
that the distribution function of $\tau$ under  $W\otimes \eta$ is exponential.
 In particular, it is absolutely continuous. A similar argument
 shows that if $Q\notin \M_A$ then
$E^{Q_n}[\s(t)]$ converges pointwise in $t$ to $E^{Q}[\s(t)]$, for
all those $t$ such that $Q(\tau=t)=0$.   \\ Taking $f(\l)\equiv 1$
we simply have that for all $t$, $ E^Q[\s(t)]$ is a continuous
mapping  in $Q$ on $\mathcal M_A$. Choosing instead $f(\l)=e^{-\g}$,
$f(\l)=-\g$ and $f(\l)=\a$, we prove continuity for $E^Q\left[ \s(t)
\
e^{-\g}\right]$, $E^Q\left[-\g \s(T)\right]$ and $m_Q(t)$ respectively.\\
The next step is to show that $Q_n\to Q$ implies that \be{EE}\left|
E^Q\!\left[\int_0^T (1-\s(t))\ e^{-\g} e^{\b m_Q(t)} dt\right] \! -
E^{Q_n}\!\left[\int_0^T (1-\s(t))\ e^{-\g} e^{\b m_{Q_n}(t)}
dt\right]   \right|  \ee converges to zero.

We  add and subtract $E^{Q_n}\left[\int_0^T (1-\s(t))\ e^{-\g}
e^{\b m_Q(t)} dt\right]$ :
$$ \left|   E^Q\left[\int_0^T (1-\s(t))\ e^{-\g}
e^{\b m_Q(t)} dt\right]   - E^{Q_n}\left[\int_0^T (1-\s(t))\
e^{-\g} e^{\b m_Q(t)} dt\right]  + \right. $$
$$+\left.  E^{Q_n}\left[\int_0^T \left( (1-\s(t))\
e^{-\g}\right)\left( e^{\b m_Q(t)} -e^{\b m_{Q_n}(t)} \right)
dt\right] \right|\leq |a_n|+|b_n|$$ where
$$a_n=  E^Q\left[\int_0^T (1-\s(t))\ e^{-\g}
e^{\b m_Q(t)} dt\right]   - E^{Q_n}\left[\int_0^T (1-\s(t))\
e^{-\g} e^{\b m_Q(t)} dt\right] ;$$
$$ b_n=  E^{Q_n}\left[\int_0^T \left( (1-\s(t))\
e^{-\g}\right)\left( e^{\b m_Q(t)} -e^{\b m_{Q_n}(t)} \right)
dt\right].  $$ $|a_n|$ goes to zero by weak convergence.

 Concerning
$b_n$ we see that
$$|b_n|\leq  \int_0^TE^{Q_n}\left[   (1-\s(t))\
e^{-\g} \cdot\left| e^{\b m_{Q}(t)} -e^{\b m_{Q_n}(t)} \right|
\right]dt\leq$$
$$\leq
\int_0^TE^{Q_n}\left[    (1-\s(t))\ e^{-\g} \cdot\left| e^{\b
m_{Q}(t)} -e^{\b m_{Q_n}(t)} \right| \right]dt .$$ We now show
that
$$
 E^{Q_n}\left[    (1-\s(t))\ e^{-\g} \cdot\left|
e^{\b m_{Q}(t)} -e^{\b m_{Q_n}(t)} \right| \right]  \to 0 .$$ We
can rewrite it as
$$
 E^{Q_n}\left[   (1-\s(t))\ e^{-\g} e^{\b m_{Q}(t)}   \cdot\left|
e^{\b\left[ m_{Q}(t)-m_{Q_n}(t)\right] } -1 \right| \right] \leq$$
$$\leq K E^{Q_n}\left[ \left|
e^{\b\left[m_{Q}(t)-m_{Q_n}(t)\right] } -1 \right| \right]=(*)$$
for a suitable $K\in \mathbb R^+$, where we have used the fact
that $   (1-\s(t))\ e^{-\g} e^{\b m_{Q}(t)}  $ is uniformly
bounded. We now look at the term into the expectation
$$ \left| e^{\b\left[m_Q(t)-m_{Q_n}(t)\right] } -1 \right| \leq K_2 |m_Q(t)-m_{Q_n}(t)|
$$
again by uniformly boundedness, for a suitable $K_2$. Thus
$$(*)\leq
K_3 |m_Q(t)-m_{Q_n}(t)|,$$ and this
converges to zero thanks to what we have shown in (\ref{canc}).\\
As a consequence,
$|b_n|$ goes to zero as well, since we are allowed to interchange the  limit and the time integral, by dominated convergence.\\

It remains to show the continuity of the term \be{last}
E^Q[\b\s(T)m_q(\tau^-)]. \ee

Indeed, take a sequence $Q_n\to Q$, $Q\in\M_A$, then
\begin{multline*}\left|E^{Q_n}[\b\s(T)m_{Q_n}(\tau^-)]-E^Q[\b\s(T)m_Q(\tau^-)]\right| \\ \leq  \left|E^{Q_n}[\b\s(T)\left\{m_{Q_n}(\tau^-)-m_{Q}(\tau^-)
\right\}]\right|+\left|E^{Q_n}[\b\s(T)m_Q(\tau^-)]-E^Q[\b\s(T)m_Q(\tau^-)]\right|.
\end{multline*} The second term goes to zero by weak convergence
since the function $m_Q$ is  continuous. Concerning the first
term, it is enough to show that $\left\{
m_{Q_n}(t)-m_{Q}(t)\right\}$ converges to zero uniformly on
$[0,T]$. To show this, we fix $t\in[0,T]$ then the following facts
hold true: \bi \item[(a)] For $\d_1>0$ there exists $\e>0$ such
that $|s-t|\leq\e\Rightarrow | m_Q(s)-m_Q(t) |\leq \d_1 $
\item[(b)] There exists $\bar n$ such that $\forall n\geq \bar n$ \\
$|m_{Q_n}(t+\e)-m_Q(t+\e)|\leq \d_2\ ;\ \ $
$|m_{Q_n}(t-\e)-m_Q(t-\e)|\leq \d_3$ \ei Point $(a)$ is due to the
continuity of $m_Q(t)$ for $Q\in\M_A$ whereas $(b)$ follows by the
fact that $ m_{Q_n}(t)\to m_{Q}(t)$ pointwise in $t$ as shown in
(\ref{canc}). Notice that when $t+\e>T$ or $t-\e<0$
the inequalities in $(b)$ are modified appropriately without loss of generality.\\
We now claim that fixing $s\in O_t:=[t-\e,\ t+\e]$ we have \be{C1}
|m_{Q_n}(s) -  m_{Q}(s)|\leq \d_1+\d_2+\d_3=\bar \d.\ee Fix $s$
and $n$ and suppose $m_{Q_n}(s)\leq m_Q(s)$ (the other case is
treated in the same way):
$$  m_Q(s)-m_{Q_n}(s) \leq m_Q(t+\e)-m_Q(t-\e)+m_Q(t-\e)-m_{Q_n}(t-\e)\leq \d_1+\d_3;$$
where we have used the fact that $m_Q(t)$ is increasing in $t$ for
all $Q\in\M_1$. Thus we  can extract a finite covering
$\{O_{t_k}\}$ of $[0,T]$ where (\ref{C1}) holds true hence uniform
convergence is proved. \epr

\emph{Proof of Theorem \ref{t1}.\\} We denote by $\PP_N$ the
distribution of $\rho_N$ under $P_N$, i.e. $\PP_N:=P^N\circ
\rho_N^{-1}$. We now state a LDP for the sequence $\{\PP_N\}_N$.
Thanks to Lemma \ref{l1}, we have identified the Radon Nikodym
derivative that relates
 $W^{\otimes N}$ and $P^{\ul}_N$
  (where $W^{\otimes N}$ plays the role of the reference measure).
  A natural way to develop a large deviation principle is now to rely on
  Lemma \ref{varad2}.

Since $(\s_i[0,T];\l_i)$ are i.i.d. random variables under
$(W\otimes \mu) ^{\otimes N}$, we can apply Sanov's Theorem (see
\cite{VA}) to the sequence of measures $(\mathcal W_N)_N$, where
$\mathcal W_N$ represents the law of the empirical measure in the
case of independence (i.e. under $(W\otimes\mu)^{\otimes N}$). Hence
$(\mathcal W_N)_N$ obeys a large
deviation principle with rate function $H(Q|W\otimes \mu)$.\\
Being $F(Q)$ bounded in the weak topology and continuous on
$\M_A\supset \M_H$ where  $\mathcal M_A=\{Q\in\M_1|Q\ll W  \otimes
 \mu\} $ and $ \mathcal M_H=\{Q\in \M_1:H(Q|W\otimes \mu)<\infty\}$, we
can rely on Lemma \ref{varad2} to conclude that the sequence
$(\mathcal P_N)_N$ obeys a large deviation principle with good rate
function
$$I(Q)=H(Q|W\otimes \mu)-F(Q)-\inf_{R\in \M_1}[H(R|W\otimes \mu)-F(R)].$$
We finish the proof by showing that \be{VarQW2} \inf_{R\in
{\M}_1}[H(R |W\otimes
\mu)-F(R)]=\lim_{N\to\infty}\frac{1}{N}\log\left[\int_{\M_1}e^{NF(Q)}\mathcal
W_N(dQ)\right]=0.\ee The first equality is
 simply a consequence of Equation
(\ref{lub4}).\\
We are thus left to prove that $\int_{ \M_1} e^{NF(Q)}\mathcal
W_N(dQ)=1$. Indeed,
$$\mathcal P_N (\,\cdot\ ) = \int \mu^{\otimes N} (d\ul) P^{\ul}_N(\rho_N\in \cdot\  )=$$
$$= \int  \mu^{\otimes N} (d\ul) \int \mathbb I_{\{\rho_N\in \cdot\ \}} \frac{dP^{\ul}_N}{dW^{\otimes N}}  d W^{\otimes N} =
\int  \mathbb I_{\{\rho_N\in \cdot\ \}}  e^{N F(\rho_N)} \
d(W^{\otimes N}\otimes \mu^{\otimes N} ) =$$
$$=\int \mathbb I_{\{ Q\in \cdot\ \}} e^{N F(Q)} \mathcal W_N(dQ).$$
Being $\mathcal P_N( {\mathcal M}_1)=1$,  the thesis follows.\fine
\subsection{Proof of Theorem \ref{t2}}
We need to  define a new process  and a technical lemma related to it. \\
We associate with any $Q\in\M_1$ the law of a time inhomogeneous
Markov process on $\{0;1\}$ which evolves according to the following
rules:
\[
\begin{array}{llc}
\s=0 \to \s=1 \quad& \mbox{with intensity}\quad& e^{-\g}e^{\b\int \a'\eta(t) Q(d\eta[0,T],d\l)}\\
\s=1 \to \s=0 \quad& \mbox{with intensity}\quad& 0
\end{array}
\]
and with $\s_i(0)=1$ for all $i=1,...,N$.\\
We denote by $P^{\l,Q}$ the law of this process and by
$P^{Q}=P^{\l,Q}\otimes \eta$. In other words, $P^{\l,Q}$ is the law
of the Markov process on $\{0;1\}$ with initial distribution
$\delta_0$ and time-dependent generator $ L^{\l,Q}_t $ defined as
\be{Ls} L^{\l,Q}_t f(s) = (1-s)e^{-\g}e^{\b\int \a'\eta(t)
Q(d\eta[0,T],d\l)}( f(1-s)-f(s)). \ee We show now an important
property of $P^{ Q}$. \bl{repr} For every $Q \in \M_1$, we have
\[
I(Q) = H(Q|P^Q).
\]
\el
\bpr We distinguish two cases:\\
\emph{\underline{Case 1}}. $Q:H(Q|W\otimes \eta)<\infty$. We have
$$I(Q)=\int \log \frac{dQ}{d(W\otimes\eta)}dQ - F(Q).$$
By Girsanov's formula for continuous time Markov chains, we obtain
\begin{multline}
\log \frac{dP^{\l,Q}}{d W }= \int_0^{T}(1-\s(t))
\left( 1- e^{-\g} e^{\b \int Q(d\eta[0,T],d\l') \a' \eta(t)} \right)dt\\
  +    \s(T) \left[ - \g + \left(\b \int \a' \eta(t^-) Q(d\eta[0,T],d\l')\right) \Big|_{t=\tau(\s[0,T])}
  \right]
   ;\end{multline}
  hence, by definition of $F$ given in (\ref{F}) we have
$$F(Q)=\int \log \frac{dP^{\l,Q}}{dW} \ dQ$$
so that \be{I3} I(Q)=\int \log \frac{dQ}{d(W\otimes\eta)}dQ-\int
\log \frac{dP^{\l,Q}}{dW} dQ=\int \log\frac{dQ}{dP^Q}dQ \ee where
the last equality follows from
$$\frac{dQ}{d(W\otimes\eta)}\frac{dW}{dP^{\l,Q}}=
\frac{dQ}{d(W\otimes\eta)} \frac{d (W\otimes\eta)}{  dP^{ Q}}=\frac{dQ}{dP^Q}.$$
Being $\int \log\frac{dQ}{dP^Q}dQ =H(Q|P^Q)$, the thesis follows.\\

\noindent \emph{\underline{Case 2}}. $Q:H(Q|W\otimes
\eta)=+\infty$. In this case $I(Q)=+\infty$. \\Thus we have to
check that $H(Q|P^Q)=+\infty$ as well. Being $$H(Q|P^Q)=\int \log
\frac{dQ}{d(W\otimes \eta)} dQ +\int \log \frac{dW}{d
P^{\l,Q}}dQ,$$ the thesis follows since  $\int \log
\frac{dQ}{d(W\otimes \eta)} dQ =+\infty$ being $ H(Q|W\otimes
\eta)=+\infty$ and since $\int \log \frac{dW}{d P^{\l,Q}}dQ=-F(Q)$
which is bounded . \epr

\emph{Proof of Theorem \ref{t2}.}\\
By properness of the relative entropy ($H(\mu|\nu) = 0 \Rightarrow
\mu = \nu$), from Lemma \ref{Ls} we have that the equation $I(Q) =
0$ is equivalent to $Q = P^Q$. Suppose $\bar Q$ is a solution of
this last equation. In this case $\bar Q=P^{\bar Q}=P^{\l,\bar
Q}\otimes \mu$, where $P^{\l,\bar Q}$ is the law of the Markov
process on $\{0;1\}$ with initial distribution $\delta _0$ and
time-dependent generator $L^{\l,\bar Q}_t$ as defined in (\ref{Ls}).
The marginals of a Markov process are solutions of the corresponding
{\em forward equation}. This leads to the fact that $\bar q_t:=
\Pi_t P^{\l,\bar Q}$, the $t-$projection of $P^{\l,\bar Q}$, is a
solution of $\dot{q}_t=\mathcal L_t^{\l,\bar Q}$ where $\mathcal
L_t^{\l,\bar Q }$ is the adjoint of $L_t^{\l,Q}$:
$$(\mathcal L^{\l,\bar Q}_t q)(x)=
 x  e^{-\g} e^{\b\int \a'\eta(t) \bar Q(d\eta[0,T],d\l')} q(-x)-
 \mathbb
(1-x) e^{-\g} e^{\b\int \a'\eta(t) \bar Q(d\eta[0,T],d\l')}q(x).$$
More specifically, when $x=0$ we have
$$(\mathcal L^{\l,\bar Q}_t q)(0)= - e^{-\g} e^{\b\int \a'\eta(t) \bar Q(d\eta[0,T],d\l')}q(0)
$$ and when $x=1$ \be{LL}(\mathcal L^{\l,\bar Q}_t q)(1)=  e^{-\g} e^{\b\int \a'\eta(t) \bar
Q(d\eta[0,T],d\l')}q(0).\ee

We now prove that $\dot{q}_t=\mathcal L^{\l,\bar Q}_t q$ admits at
most one solution for each initial condition. To see this, define
$\bar q_t(\l):= P^{\l,\bar Q}(\s(t)=1)$. Then $ \frac{d\bar
q_t(\l)}{dt}=G(\bar q_t(\l))$, where $G(q)=e^{-\g}e^{\b\int \a'
q(\l') \mu(d\l')} (1-q(\l))$. Notice that $\bar{q}_t(\l)\in
L^1(\mu)$ and  $G(\cdot)$ is a locally Lipschitz operator on a
Banach space. Thus  $ \frac{d\bar q_t(\l)}{dt}=G(\bar q_t(\l))$ has
at most one solution in $[0,T]$,
for a given initial condition (see \cite{Brezis}, Theorem VII.3).\\
Since $P^{\bar Q}$ is totally determined by the flow $\bar q_t$,
it follows that equation $Q = P^Q$ has at most one solution. The
existence of a solution follows from the fact that $I(Q)$ is the
rate function of a LDP, and therefore {\em must} have at least one
zero.  By what shown in (\ref{LL}), $\bar{q}_t(\l)$ solves
(\ref{MKV}). Hence $Q_*$ turns out to be the unique solution of
the fixed point argument $Q = P^Q$. Moreover, it satisfies all the
conditions of Theorem \ref{t2}. \fine

\subsection{Proof of Theorem \ref{t3}}

The key technical tool for the proof of Theorem \ref{t3} is the
following result due to Bolthausen \cite{B}.

\bt{TeoB} Let $(B, \|\cdot\|)$ be a real separable Banach space. Let
$(Z_{k})_{k\geq 1}$ be a sequence of $B$-valued, i.i.d. random
variables, defined on the probability space $(\Omega,\mathcal A,
\mathbb P)$ and denote by $w$ their common law. Define
$X_N:=\frac{1}{N}\sum_{k=1}^N Z_k$ and consider a continuous map
$\Psi:B\to \mathbb R$. Suppose that the following conditions are
satisfied:
\begin{itemize}
\item[(B.1)] $\int \exp (r|x|)w(dx)<\infty$ \ \ \mbox{for all }$r\in\mathbb R$.
\item[(B.2)] For any $x\in B$, $\Psi(x)\leq C_1+C_2\|x\|$, for some $C_1,C_2>0$. Moreover, $\Psi$ is three times continuously Fr\'echet differentiable.
\item[(B.3)] Define, for $h\in B'$ (the topological dual of $B$), $\Lambda (h):=\int e^{h(z)}w(dz)$, and for $x\in B$, $\Lambda^*(x):=\sup_{h\in B'}[h(x)-\Lambda(h)]$. Assume that there exists a unique $z^*\in B$ such that $\Lambda^*(z^*)-\Psi(z^*)=\inf_{z\in B}[\Lambda^*(z)-\Psi(z)]$.
\item[(B.4)] Define the  probability $p$ on $B$ by   $\frac{dp}{dw} =  \frac{e^{ D\Psi(z_*) }}{c}$  for a suitable normalizing factor $c$.
This probability is well defined and $\int z p(dz)=z^*$. Let $p_*$
denote the centered version of $p$, i.e., $p_*=p \circ \theta
_{x^*}^{-1}$, where $\theta_a:B\to B$ is defined by
$\theta_a(x)=x-a$. For $h \in B'$ define $\tilde h \in B$ by $\tilde
h = \int z h(z) p_*(dz)$. Then we assume that for every $h \in B'$
such that $\tilde h \ne 0$
    $$\int h^2(z)p_*(dz) - D^2\Psi(z^*)[\tilde h,\tilde h]>0.$$
\item[(B.5)] $B$ is a Banach space of type 2.
\footnote{A Banach space $B$ is said to be of type $2$ if $\ell^2(B)\subseteq C(B)$.
Here $\ell^2(B)=\{(x_n)\in B^{\infty}:\sum_i \|x_i\| ^2 <\infty\}$ and
$C(B)=\{(x_n)\in B^\infty:\sum_j \e_jx_j \ \mbox{converges in probability}\}$ where
$(\e_n)$ is a Bernoulli sequence, i.e., a sequence of independent random variables such
 that $P(\e_n=\pm 1)=\frac 12$. For more details see  \cite{B}. }
\end{itemize}
Now, letting  $\pi_N$ be the probability on $(\Omega,\mathcal A)$ given by
 \be{B.4}  \frac{d\pi_N}{d\, \mathbb P} =
 \frac{e^{N \left(\Psi ( X_N)+\frac{\Sigma(X_N)}{N}\right)}}
 {E^{\mathbb P} \left[e^{N \left(\Psi ( X_N)+\frac{\Sigma(X_N)}{N}\right)}\right]}\ ,\ee
where $\Sigma$ is  linear, continuous and bounded on the support of
the law of $X_N$, uniformly in $N$. Then, for every $h_1,...,h_n\in
B'$, the  $\pi_N$-law of the $n-$dimensional vector
$$\sqrt{N} \left( h_i(X_N) -h_i (z_*)\right)_{i=1}^n$$
converges weakly, as $N\to\infty$, to the law of  a centered
Gaussian vector with covariance matrix $\mathcal C \in \mathbb
R^{n\times n}$,    such that  for $i,j=1,...,n$ \be{gulp}(\mathcal
C)_{i,j}=\int h_i(z) h_j(z) p_*(dz) -D^2\Psi(z_*)[\tilde h_i,\tilde
h_j]. \ee \et

\br{RTeoB}
\bi
\item[i.]
The Theorem in \cite{B} is stated for $\Sigma = 0$. The same proof applies with our assumptions on $\Sigma$ without changes. It is likely that these assumptions can be weakened considerably.
\item[ii.]
The Theorem in \cite{B} contains a stronger statement that the one
given here. Indeed, it is shown that the field $\sqrt{N} \left(
h(X_N) -h(z_*)\right)_{h \in B'}$ converges weakly to a Gaussian
field, while we only stated convergence for finite dimensional
distributions. This is all we need to prove Theorem \ref{t3}. Our
proof has not allowed to take advantage of the full strength of the
Theorem in \cite{B}. \ei \er

The ``natural'' space for the Central Limit Theorem in Theorem \ref{t3} is the set $\M$ of signed measures, which is not a Banach space. To apply Theorem \ref{TeoB}, we need to map $\M$ to a Banach space of type 2.
\bl{lCLT1}
The following properties hold true under the reciprocity condition (R):
\begin{itemize}
\item[i)] There exists a Banach space of type $2$ $(B,\| \cdot \|)$, a linear
map  $T:\mathcal M \to B$,
continuous on the set $\{Q:Q(\tau=T)=0\}$. Moreover there exist two
 continuous maps $\Psi,\Sigma:B\to \mathbb R$, where $\Psi$ is bounded and three times Fr\'echet
differentiable and $\Sigma$ is linear, such that
\be{dPdQ}\frac{dP^{\l}_N}{dW^{\otimes
N}} =\exp\left\{N\left[\Psi(T(\rho_N ))+\frac{\Sigma(T(\rho_N ))}{N}\right]\right\}\ ,\qquad a.s.
\ee
\item[ii)] For any vector $\Phi=(\Phi_1,...,\Phi_n)\in {\mathcal C_b}^n$ there exist
$h=(h_1,...,h_n)\subset B'$ such that
$(h_i\circ T)(Q)=\int\Phi_i dQ$, where $B'$ stands for the
topological dual of $B$.
\end{itemize}
\el \bpr The first step consists in giving an alternative expression
for  $F(Q)$ given in (\ref{F}). Look at the term
\[
 \int Q(d\s[0,T],d\l) \left\{ \s(T) \left(\b \int Q(d\eta[0,T],d\l') \a' \eta(t^-)\right) \Big|_{t=\tau(\s[0,T])}
\right\}.
\]
using the reciprocity condition (R), we obtain
\begin{multline*} b
\int Q(d\s[0,T],d\l) \left\{ \s(T) \a \int Q(d\eta[0,T],d\l')\a'
\eta(t^-) \Big|_{t = \tau(\s[0,T])} \right\}\\ = b \int
Q(d\s[0,T],d\l) Q(d\eta[0,T],d\l') \a \a' {\bf 1}_{\{\tau(\s[0,T])
\leq T\}} {\bf 1}_{\{\tau(\eta[0,T]) < \tau(\s[0,T])\}}=(*)
\end{multline*}
Note that ${\bf 1}_{\{\tau(\eta[0,T]) < \tau(\s[0,T])\}}= {\bf
1}_{\{\tau(\eta[0,T]) \leq T\}}$ unless $\tau(\s[0,T]) \leq
\tau(\eta[0,T])\leq T$. Thus \begin{multline*}(*) = b \int
Q(d\s[0,T],d\l) Q(d\eta[0,T],d\l') \a \a' {\bf 1}_{\{\tau(\s[0,T])
\leq T\}} {\bf 1}_{\{\tau(\eta[0,T]) \leq T\}}  \\ - b \int
Q(d\s[0,T],d\l) Q(d\eta[0,T],d\l') \a \a'  {\bf
1}_{\{\tau(\eta[0,T]) \leq T\}} {\bf 1}_{\{\tau(\s[0,T]) \leq
\tau(\eta[0,T])\}} \\ = b \int Q(d\s[0,T],d\l) Q(d\eta[0,T],d\l') \a
\a' {\bf 1}_{\{\tau(\s[0,T]) \leq T\}} {\bf 1}_{\{\tau(\eta[0,T])
\leq T\}}  \\  - b \int Q(d\s[0,T],d\l) Q(d\eta[0,T],d\l') \a \a'
{\bf 1}_{\{\tau(\eta[0,T]) \leq T\}} {\bf 1}_{\{\tau(\s[0,T]) <
\tau(\eta[0,T])\}} \\ - b  \int Q(d\s[0,T],d\l) Q(d\eta[0,T],d\l')
\a \a'  {\bf 1}_{\{\tau(\eta[0,T]) \leq T\}} {\bf
1}_{\{\tau(\s[0,T]) =\tau(\eta[0,T])\}}.
\end{multline*}
Therefore
\begin{multline*}
b \int Q(d\s[0,T],d\l) \left\{ \s(T) \a \int Q(d\eta[0,T],d\l')\a'
\eta(t^-) \Big|_{t = \tau(\s[0,T])} \right\}\\ = \frac{b}{2}  \int
Q(d\s[0,T],d\l) Q(d\eta[0,T],d\l') \a \a' {\bf 1}_{\{\tau(\s[0,T])
\leq T\}} {\bf 1}_{\{\tau(\eta[0,T]) \leq T\}} \\ - \frac{b}{2} \int
Q(d\s[0,T],d\l) Q(d\eta[0,T],d\l') \a \a'  {\bf
1}_{\{\tau(\eta[0,T]) \leq T\}} {\bf 1}_{\{\tau(\s[0,T])
=\tau(\eta[0,T])\}} \\ = \frac{b}{2} \left[\int Q(d\s[0,T],d\l)
\a{\bf 1}_{\{\tau(\s[0,T]) \leq T\}}\right]^2 - \frac{b}{2} \sum_{t
\in [0,T]} \left[ \int Q(d\s[0,T],d\l) \a {\bf 1}_{\{\tau(\s[0,T]) =
t \}} \right]^2.
\end{multline*}

\vskip 1cm

Thus, defining
\begin{multline} \label{CLT4}
F_1(Q) :=
\int Q(d\s[0,T],d\l) \left\{ \int_0^T (1-\s(t))
\left( 1- e^{-\g} e^{\b \int Q(d\eta[0,T],d\l') \a' \eta(t)} \right)
dt  - \g \s(T)  \right\} \\ + \frac{b}{2} \left[\int Q(d\s[0,T],d\l) \a \s(T) \right]^2
\addtocounter{for}{1}
\end{multline}
and
\be{CLT5} F_2(Q) := - \frac{b}{2} \sum_{t \in [0,T]} \left[ \int
Q(d\s[0,T],d\l) \a {\bf 1}_{\{\tau(\s[0,T]) = t \}} \right]^2,\ee
we have that 
$F( Q) = F_1( Q) + F_2(Q)$. Lemma \ref{l1} thus holds also after
replacing $F$ by $F_1 + F_2$. Let $M>0$ be a constant such that,
under $\eta$, the random parameters $\a$ and $\g$ have absolute
value less that $M/2$. Now we define the following maps:
\[
T_1 : \begin{array}{rcl} \M & \ra & L^2[0,T] \\ Q & \mapsto & \frac{1}{2}\int Q(d\s[0,T],d\l) [1-\s(t)] \end{array}
\]
\[
T_2 : \begin{array}{rcl} \M & \ra & L^2([0,T]\times \N) \\ Q & \mapsto & C(M) \int Q(d\s[0,T],d\l)  \left[ e^{-\g} \frac{(M\b)^n}{n!} (1-\s(t)) \right] \end{array}
\]
\[
T_3 : \begin{array}{rcl} \M & \ra & L^2[0,T] \\ Q & \mapsto & \int Q(d\s[0,T],d\l) \frac{\a}{M} \s(t) \end{array}
\]
\[
T_4 : \begin{array}{rcl} \M & \ra & \R\\ Q & \mapsto & \int Q(d\s[0,T],d\l) \frac{\g}{M} \s(T) \end{array}
\]
\[
T_5 : \begin{array}{rcl} \M & \ra & \R\\ Q & \mapsto & \int Q(d\s[0,T],d\l) \frac{\a}{M} \s(T) \end{array}
\]
\[
T_6 : \begin{array}{rcl} \M & \ra & \R\\ Q & \mapsto & \int Q(d\s[0,T],d\l) \a^2 \s(T) \end{array},
\]
where $C(M)$ is some positive constant such that $C(M) e^{-\g} e^{M\b} \leq \frac{1}{2}$ $\eta$ almost surely.
Note that, for $Q \in \M_1$ and $i=1,2,\ldots,5$, we have that $|T_i(Q)| \leq \frac{1}{2}$.
Now, let $g: \R \ra \R$ be a ${\cal{C}}^{\infty}$ function such that $g(x) = x$
for $|x| \leq 1/2$,  $g(x) = 0$ for $|x| > 3/4$ and $\| g\|_{\infty} \leq 3/4$. For
\[
z = (z_1,z_2,z_3,z_4,z_5,z_6) \in L^2[0,T]  \times  L^2([0,T]\times
\N) \times L^2[0,T] \times R \times R \times R
\]
we set \be{CLT6} \Psi(z) := 2 \int_0^T g(z_1(t))dt -
\frac{1}{C(M)} \sum_n \int_0^T g((z_2(t,n)) g(z_3(t))^n dt -
Mg(z_4) + \frac{b M^2}{2} g^2(z_5) \ee and \be{CLT7} \Sigma(z) :=
-\frac{b}{2} z_6. \ee We now claim that, for $Q \in \M_1$ and
setting $T = (T_1,T_2,T_3,T_4,T_5,T_6)$, \be{F1Psi} F_1(Q)=
\Psi(T(Q)). \ee Moreover \be{F2S}F_2(\rho_N)=
\frac{\Sigma(T(\rho_N))}{N},\quad W\otimes\mu-a.s.\ee so that
(\ref{dPdQ}) holds. Equation (\ref{F1Psi})is straightforward;
(\ref{F2S}) follows since
$$F_2(\rho_N)=-\frac{b}{2}\sum_t\left(\frac{1}{N}\sum_i \a_i
\Delta \s_i(t)\right)^2=-\frac{b}{2}\frac{1}{N^2} \sum_i\a_i^2\sum_t
(\Delta \s_i(t))^2=-\frac{b}{2}\frac{1}{N^2} \sum_i\a_i^2\s_i(T),$$
where the one to last equality follows since simultaneous jumps may
happen only with zero $(W\otimes \mu)^{\otimes N}-$probability. We
thus have that $F_2(\rho_N)=-\frac{b}{2}\frac{1}{N}\int \a^2\s(T)
d\rho_N$ and the claim follows by definition of $\Sigma$.

Now set
\[
B := L^2[0,T]  \times  L^2([0,T]\times \N) \times L^2[0,T] \times \R
\times \R \times \R \times \R^n.
\]
Clearly $B$ is a Hilbert space (hence a Banach space of type 2), and
the maps $\Psi,\Sigma$ are trivially extended to $B$. Moreover, the
map $T$ can be completed to a $B$-valued map by letting, for
$i=1,2,\ldots,n$,
\[
T_{6+i}(Q) := \int \Phi_i dQ,
\]
where $\Phi=(\Phi_1,...,\Phi_n) \in {\mathcal C_b}^n$ is given.

To complete the proof of part i) of Lemma \ref{lCLT1} one has to show the desired regularity of $\Psi$ and $\Sigma$. The only nontrivial fact is to show regularity of the term
\[
\sum_n \int_0^T g((z_2(t,n)) g(z_3(t))^n dt.
\]
However, the fact that $\| g\|_{\infty} \leq 3/4$ allows to control
the tails of the sum above; continuity and Fr\'echet
differentiability of any order is obtained by standard estimates,
The details are omitted.

Finally, to prove part ii), for $B \ni (z_1,\ldots,z_6,
\ldots,z_{6+n})$, it is enough to define for $i=1,2,\ldots,n$
\[
h_i (z) = z_{6+i}.
\]
\epr

\emph{Proof of Theorem \ref{t3}.}\\ Having identified a suitable
Banach space, Theorem \ref{t3} immediately follows from Theorem
\ref{TeoB} applied to the sequence $Z_i=T(\delta_{\{\s_i[0,T],\l
\}})$ taking values on $(B,\|\cdot \|)$. Notice that in our setting
$\Omega=(\D[0,T]\times \R^2)^N$ and $\P=(W\otimes \mu)^N$. Theorem
\ref{t3} is guaranteed by  the following three facts:
\begin{enumerate}
\item $P_N\equiv \pi_N$, where $\pi_N$ is the probability
appearing in Theorem \ref{TeoB}; \item $\left( \int \Phi_i d
\rho_N -\int \Phi_i dQ_* \right)_{i=1}^n=\left( h_i(X_N)-h_i(z_*)
\right)_{i=1}^n$; \item $ \int
(\phi_i-\phi_i^*)(\phi_j-\phi_j^*)dQ_*-D^2F(Q_*)[\hat\Phi_i,\hat\Phi_j]=\int
h_i(z) h_j(z) p_*(dz) - D^2\Psi(z_*)[\tilde h_i,\tilde h_j]$.
\end{enumerate}
Point $1.$ follows from the definition of $\pi_N$ and from
equations (\ref{F1Psi}) and (\ref{F2S}). Point $2.$ is a
consequence of the fact that $z_*=T(Q_*)$ and $h_i\circ
T(Q_*)=\int \Phi_i dQ_*$. Point $3.$ will be proved in details in
Lemma \ref{L.A.3.2} (see in particular equation (\ref{C=C})). An
immediate application of equations (\ref{C=C}) and (\ref{DeltaA})
finally guarantees the validity of (\ref{CovDiag}). \\
Assuming point $3.$, it remains to show the validity of the
central limit theorem in $B$. In other words we need to check the
five assumptions of Theorem \ref{TeoB}. $(B.1)$, $(B.2)$ and
$(B.5)$ are easy to see. $(B.3)$ and $(B.4)$ are not
straightforward. The rest of this section is devoted to the proof
that these two assumptions are satisfied.
\\
We begin to prove $(B.3)$. We define two sequences
of measures on $B$  as follows:
$$p_N(\cdot)=\mathcal P_N\circ T^{-1}(\cdot)\qquad w_N(\cdot)=\mathcal W_N\circ
T^{-1}(\cdot).$$ From (\ref{dPdQ}) it can be shown that \be{CLT8}
\frac{dp_N}{dw_N}=e^{N\left(\Psi+\frac{\Sigma}{N}\right)}\ee for
$\Psi$ and $\Sigma$ as defined in (\ref{CLT6}) and (\ref{CLT7}).\\
By the contraction principle (see Theorem 4.2.1 in  \cite{DZ}),
the sequence $(p_N)_N$ satisfies a LDP with the good rate function
$J(z)=\inf_{Q=T^{-1}(z)}I(Q).$ Being $Q_*$ the unique zero for
$I$, $J$ has a unique zero $z_*=T(Q_*)$.

A LDP for the sequence $(p_N)_N$ can be obtained in an alternative way. Indeed, we notice that
$w_N$ is the law of the
random variables
$$X_N =\frac{1}{N} \sum_{i=1}^N Z_i\ \in B,$$
where $Z_i$ are i.i.d. $B-$valued random variables with law $w$.
Thus  we have that $(w_N)_N$ satisfies a (weak) LDP with rate function
$\Lambda^*$, with $\Lambda^* (z) :=
\sup _{\varphi\in B'} \left\{ \varphi(z) - \Lambda
(\varphi)\right\}$ and $ \Lambda(\varphi):=\ln \int
e^{\varphi(z)}\ w(dz)$. Thus, applying Varadhan's Lemma, $(p_N)_N$ satisfies a (weak) LDP with rate function
$\Lambda^*(z)-\Psi(z)$. Since the rate function is unique, it follows that
\[
J(z) = \Lambda^*(z)-\Psi(z).
\]
Having proved already that $J(z)$ has a unique zero, the proof of (B3) is completed.

We are thus left to show $(B.4)$: for each $\lambda\in B'$ such that
$\tilde h = \int zh(z) p_*(dz)\ne 0$ we have \be{varL}\int
h^2(z)p_*(dz) - D^2\Psi(z^*)[\tilde h,\tilde h]>0;\ee
where $p$ and $p_*$ are defined in Theorem \ref{TeoB}.\\
This proof is rather technical and long. We divide it into three
steps. We first show that the measure $p$ such that
$\frac{dp}{dw}= {e^{D\Psi(z_*)}} $   is exactly the law of the
random variable $T(\delta_{\{\s[0,T],\l\}})$ induced by $Q_*$.
This argument is then used in the second step to ensure the
positivity of a suitable functional $\mathcal H:\mathcal C_b\times
\mathcal C_b \to \mathbb R$.
In the last part we see how to relate $\mathcal H$ to assumption $(B.4)$.\\

\noindent \emph{\underline{Step 1:}}  The key result of this first
step is given in Lemma \ref{lCLT2} below. We look at the measure
$p$ on $B$, defined by
$$\frac{dp}{dw}(z)=e^{D\Psi(z_*)[z]}\ ,\quad \mbox{with} \ z_*=T(Q_*)$$
where, as already seen, $w$ represents the law of
$T(\delta_{\{\s[0,T],\l\}})$ induced by $W\otimes \mu$.\\
We shall prove in Lemma \ref{lCLT2} that $p$ is the law of
$T(\delta_{\{\s[0,T],\l\}})$ induced by $Q_*$. \bl{lCLT2} The
measure $p$ is the law of $T(\delta_{\{\s[0,T],\l\}})$ induced by
$Q_*$. \el \bpr We first prove the following two claims \bi\item[i)]
\begin{eqnarray}\label{Fclaim} DF(Q_*)[\delta_{\{\s[0,T],\l\}}]=\log
\frac{dQ_*}{d(W\otimes\mu)}(\s[0,T],\l), \end{eqnarray} for
$W\otimes \mu-$almost all $(\s[0,T],\l)$. \item[ii)]
\be{Fclaim2}DF_2(Q)[r]=0\ee for all $Q\in\M_1$ such that $\int \a'
1_{\{\tau(\s[0,T])=t\}} dQ =0$, where $F_2$  is defined in
(\ref{CLT5}).\ei To prove the claim we need to compute
$DF(Q_*)[\delta_{\{\s[0,T],\l\}}]$, i.e. the Fr\'echet derivative of
the function $F$ at $Q_*$ in the direction
$\delta_{\{\s[0,T],\l\}}$. An explicit computation reveals that for
$Q\in\M_1$ and $r\in\M$, $DF(Q)[r]$ is well defined and in
particular
\begin{multline*}    DF(Q)[r]=\lim _ {h\to 0} \frac {F(Q+h r) - F(Q)}{h} =\\
  =  \int dQ \int_0^{T}(1-\s(t)) m_r(t)  \ e^{-\g+\b m_Q(t)} dt +
  \int dr \int_0^{T}(1-\s(t))   (1-e^{-\g+\b m_Q(t) })dt+\nonumber \\
  +  \int d r \ \left[-\s(T)
(\g-\b m_Q(\tau^-))\right] + \int dQ  \ \left[\s(T) m_r(\tau^-)
\right],
\end{multline*}
where we have put as usual $m_p(t)=\int \a'\eta(t)
p(d\eta[0,T],d\a')$ for $p\in\M$, $t\in[0,T]$.  \\We now compute
$DF(Q_*)[\delta_{\{\s[0,T],\l\}}]$. Notice that
$m_{\delta_{\{\s[0,T],\l\}}}(t)=\a\s(t)=0$ for all
$t<\tau(\s[0,T]) $ so that \be{DFQ*}
DF(Q_*)[\delta_{\{\s[0,T],\l\}}]
  =   \int_0^{T}(1-\s(t))   (1-e^{-\g+\b m_{Q_*}(t) })dt-
      \left[ \s(T)
(\g-\b m_{Q_*}(\tau^-))\right]. \ee By virtue of Girsanov's Formula
for Markov chains it can be seen that
$$\int_0^{T}(1-\s(t))   (1-e^{-\g+\b m_{Q_*}(t) })dt-
      \left[ \s(T)
(\g-\b m_{Q_*}(\tau^-))\right]= \log \frac{dP^{Q_*}}{d(W\otimes
\mu)}
$$
where $P^Q$ is the law of the Markov process with generator given in
(\ref{Ls}). (\ref{Fclaim}) thus follows since $P^{Q_*}=Q_*$ as shown
in the proof of Theorem \ref{t2}. Concerning $(ii)$, notice that
\begin{multline*} F_2(Q + h r)=-\frac{b}{2}\sum_t\left(\int \a'
 {\bf{1}}_{\{\tau(\s[0,T])=t\}} d\{Q+hr\}\right)^2\\
=-\frac{b}{2}\sum_t \left(\int \a' {\bf{1}}_{\{\tau(\s[0,T])=t\}}
dQ \right)^2 - b h \sum_t\left( \int \a'
{\bf{1}}_{\{\tau(\s[0,T])=t\}} dQ \int \a'
 {\bf{1}}_{\{\tau(\s[0,T])=t\}} dr   \right) \\-\frac{b}{2} h^2 \sum_t\left(\int \a'
 {\bf{1}}_{\{\tau(\s[0,T])=t\}} dr \right)^2.
\end{multline*}
When $\int\a'  {\bf{1}}_{\{\tau(\s[0,T])=t\}} dQ=0$, the first two
terms of the last expression vanish. Moreover under the same
hypothesis $F_2(Q)=0$. Hence \be{F2=0}DF_2(Q)[r]=\lim_{h\to 0}
\frac{1}{h} \left[F_2(Q + h r) - F_2(Q)\right]=\lim_{h\to 0}
-\frac{b}{2} \frac{ h^2}{h} \sum_t\left(\int \a'
 {\bf{1}}_{\{\tau(\s[0,T])=t\}} dr \right)^2 =0.\ee Notice that
in writing the latter equality we have implicitly used the fact
that\\ $ \sum_t\left(\int \a' {\bf{1}}_{\{\tau(\s[0,T])=t\}} dr
\right)^2<\infty$. This is true since for any $r\in\M$
$$0\leq  \sum_t\left(\int \a' {\bf{1}}_{\{\tau(\s[0,T])=t\}} dr
\right)^2 \leq \|\, \a\|^2\cdot  |\, r|^2_{TV}<\infty$$ where
$\|\a\|$ stands for the supremum of $\a$ in the support of
$\mu$ and $|\, r|_{TV}$ denotes the \emph{total variation} of $r$. \\
As a corollary of claim $(ii)$ above we see that
$$D\Psi(T(Q_*))[T(r)]=\lim_{h\to 0}\frac{\Psi(T(Q_*+hr))-\Psi(T(Q_*))}{h}=$$$$ =\lim_{h\to 0}\frac{F_1(Q_*+hr)-
F_1(Q_*)}{h}=DF_1(Q_*)[r]=DF(Q_*)[r].$$ Here we have used
(\ref{F2=0}), the fact that  $\int\a'
{\bf{1}}_{\{\tau(\s[0,T])=t\}} dQ_*=0$ since $Q_*\ll W\otimes \mu$
and equation  (\ref{F1Psi}).

Back to the statement of the lemma, we see that for $h$ measurable
and bounded
$$\int h(z) p(dz) =\int h(z)  {e^{D\Psi(z_*)[z]}} w(dz) = $$
 $$=
 \int h(T(\delta_{\{\s[0,T],\l\}}))
  {e^{DF(Q_*)[\delta_{\{\s[0,T],\l\}}]}}   \ (W\otimes \mu)  (d\s[0,T],d\l)   =
 \int h(T(\delta_{\{\s[0,T],\l\}})) dQ_* $$
where in the last  equality we have used  (\ref{Fclaim}).\epr

\noindent \emph{\underline{Step 2:}} The key result of the second
step is given in Lemma \ref{Prop1} below. It involves the measures
$\hat\Phi$ and $\hat\Phi^*$ defined in (\ref{Hp}). First of all,
it is not difficult to show that $\hat \Phi$ is absolutely
continuous w.r.t. $Q_*$ and in particular
\be{radon_3}\frac{d\hat\Phi}{dQ_*}=\Phi-\Phi^*.\ee Indeed, observe
that,  given  $\hat\Phi$  as in  (\ref{Hp}), we have
$$\hat\Phi(S)=\int_{\mathcal M_0}\left[ R(S)  \int \Phi dR\right] \nu_*(dR)=$$
$$=\int_{\D[0,T]\times Supp(\mu)}  ( {\bf{1}}_{\{(\s[0,T],\l)\in S\}} -Q_*(S))\cdot
\left(\int \Phi \ d\delta_{\{\s[0,T],\l \}}-\int \Phi dQ_*
\right)dQ_*=$$
$$=\int_{ \D[0,T]\times Supp(\mu) } \left[ ( {\bf{1}}_{\{(\s[0,T],\l)\in S\}} -Q_*(S)) \cdot (\Phi(\s[0,T],\l)-\Phi^*) \right]dQ_* $$
for any $S\subset  \D[0,T]\times Supp(\mu)$. The second equality
follows since $\nu_*$ is the law of the random variable
$\delta_{\{\s[0,T],\l\}}-Q_*$ induced by $Q_*$.
\\Notice  that
$Q_*(S) \int _{ \D[0,T]\times Supp(\mu)}   (\Phi-\Phi^*) dQ_*=0$,
being $\Phi ^*$ the expectation under $Q_*$ of $\Phi(\cdot )$. Hence
$$\hat\Phi(S)=\int_{ S }      (\Phi  -\Phi^*) dQ_*  $$
 and (\ref{radon_3}) follows.
\bl{Prop1} Given $\Phi_1$ and $\Phi_2$ in $\mathcal C_b$, let
\be{D3A}\mathcal
H(\Phi_1,\Phi_2):=Cov_{Q_*}(\Phi_1,\Phi_2)-D^2F(Q_*)[\hat\Phi_1,\hat\Phi_2];\ee
where $Cov_{Q_*}(\Phi_1,\Phi_2):=\int
(\Phi_1-\Phi^*_1)(\Phi_2-\Phi^*_2)dQ_*$. Then
 $$\mathcal H(\Phi,\Phi)>0,\ \mbox{for\ all}\ \Phi\ \mbox{such\ that}\ \hat\Phi \neq 0.$$
\el \bpr A tedious but straightforward computations provides the
second order derivative of $F$:
\begin{multline*}D^2F(Q_*)[\hat\Phi_i,\hat\Phi_j]
 = E^{Q_*} \left[\int_0^{T }-(1-\s(t))
 \beta^2 m _{\hat\Phi_i}(t)  m _{\hat\Phi_j}(t) e^{-\g+\b m _{Q_*}(t)} dt \right.  \\
 -  (\Phi_j-\Phi_j^*)\!  \int_0^{T}(1-\s(t))
\beta  m _{\hat\Phi_i}(t)  e^{-\g+\b m _{Q_*}(t)} dt \\ - \left.
   (\Phi_i-\Phi^*_i) \!\int_0^{T}(1-\s(t))   \beta
 m _{\hat\Phi_j}(t)   e^{-\g+\b m_{Q_*}(t)}  dt  \right ] \\
 +  E^{\hat\Phi_i} \left[ \s(T)  \beta  m _{\hat\Phi_j} (\tau^-)  \right]+ E^{\hat\Phi_j}
\left[\s(T)\b    m _{\hat\Phi_i}(\tau^-) \right] . \end{multline*}
Notice that we have written $\b$ instead of $b\alpha$: the
reciprocity condition is not necessary in this calculation. We now
show that $\mathcal H(\Phi,\Phi)$ is the expected value of a
square. Indeed
\begin{multline*}
 \mathcal H(\Phi ,\Phi )=Cov_{Q_*}(\Phi ,\Phi )-D^2F(Q_*)[\hat\Phi ,\hat\Phi ]=\\
  =   E^{Q_*}[(\Phi -\Phi ^*)^2 ]+ E^{Q_*}
\left[\int_0^{T}(1-\s(t))  \beta^2 [m _{\hat\Phi}(t)]^2
e^{-\g+\b m_{Q_*}(t)} dt \right.   \\
  +    \left.
   2(\Phi -\Phi^* ) \int_0^{T} (1-\s(t))  \beta
 m _{\hat\Phi} (t)  e^{-\g+\b m_{Q_*}(t)}  dt  \right] -
   2 E^{\hat\Phi } \left[ \s(T)\beta    m _{\hat\Phi} (\tau^-)  \right]
        .
\end{multline*}
The latter expectation  can be rewritten as
$$E^{\hat\Phi } \left[ \s(T)\beta    m _{\hat\Phi} (\tau^-)
\right]=E^{Q_*}\left[ (\Phi-\Phi^*) \int_0^T (1-\s(t))\b
m_{\hat\Phi}(t^-) dN(t) \right]$$ where we have used (\ref{radon_3})
and where  
  $(N(t))_{t\in[0,T]}$   defined by $N(t):=\mathbb
{\bf{1}}_{\{\tau\geq t\}}$,
is the Poisson process with intensity $\int_0^t (1-\s(s))e^{-\g+\b
m_{Q_*}(s)}ds$. \\ Recall  that $M(t) =N(t)-\int_0^t (1-\s(s))
e^{-\g+\b m_{Q_*}(s)} ds$ defined in (\ref{Mart}) is nothing but its
\emph{compensated} $Q_*-$martingale. Hence $$\mathcal H(\Phi ,\Phi
)= E^{Q_*}[(\Phi -\Phi ^*)^2 ]+ E^{Q_*} \left[\int_0^{T}(1-\s(t))
\beta^2 [m _{\hat\Phi}(t)]^2 e^{-\g+\b m_{Q_*}(t)} dt \right] $$
$$-E^{Q_*}\left[2(\Phi-\Phi^*) \int_0^T (1-\s(t))\b m_{\hat\Phi}(t)dM(t) \right]$$
By the isometry property of square integrable martingales (and
relying on the same argument used to prove (\ref{isom})), we have
$$E^{Q_*} \left[\int_0^{T } (1-\s(t)) \beta^2 m _{\hat\Phi} (t) ^2
e^{-\g+\b m_{Q_*}(t)} dt \right]= E^{Q_*} \left[ \ \left( \int_0^{T
} (1-\s(t)) \beta  m _{\hat\Phi }(t) dM(t) \right)^2\ \right].$$


Hence \be{DeltaA}
 \mathcal H(\Phi ,\Phi ) =
    E^{Q_*}\left[      \left(   (\Phi -\Phi ^*) -    \int_0^{T }  (1-\s(t)) \beta
 m _{\hat\Phi} (t)     dM(t)         \right)^2     \right]
        .
\ee
 $\mathcal H(\Phi,\Phi)$ is thus the expected value of a square, hence it cannot be negative.
 For this reason,  we simply need to  prove that it is non-zero.
 Without loss of generality we take $\Phi^*=0$.
Suppose by way of contradiction that $\mathcal H(\Phi,\Phi)=0$. Then
necessarily
$$ \left(    \Phi  (\s[0,T],\l ) -    \int_0^{T}(1-\s(t))   \beta
 m _{\hat\Phi} (s)     dM(s)         \right)=0,\quad Q_* \ a.s.$$
 Using the fact that $$m_{\hat\Phi}(s)=\int \a \s(s) \   \hat\Phi
 (d\s[0,T],d\l)=\int\a \s(s) \  \Phi(\s[0,T],\l)  \ Q_*(d\s[0,T],d\l), $$
 where the last equality follows since $\frac{d\hat\Phi}{dQ_*}=\Phi$. We
 rewrite the expression above as
\be{Pcond2} \Phi(\s[0,T],\l)=\int_0^{T}(1-\s(t)) \beta \left[ \int
\a\s(s) \Phi(\s[0,T],\l) dQ_* \right]dM(s)\ ,\quad Q_*-a.s.\ee On
the other hand, define $\Phi_t = E^{Q_*}[\Phi|\mathcal F_t]$,
where
$$\mathcal F_t=\sigma\{\s_s:0\leq s\leq t ;\  \l \}.$$
 Notice that $$\int \s(t) \Phi(\cdot) dQ_* =E^{Q_*}[\a\s(t)\Phi(\cdot) ]=
 E^{Q_*}[\a\s(t)E^{Q_*}[\Phi(\cdot)|\mathcal F_t ] ]= \int \a\s(t) \Phi_t dQ_*. $$
Taking the conditional expectation in (\ref{Pcond2}), we obtain
\begin{eqnarray*}
\Phi_t &=&E^{Q_*}\left[\left.\int_0^{T }(1-\s(t))\beta  \left( \int
\a\s(s) \Phi_s dQ_* \right)dM(s)\right|\mathcal F_t\right]
,\quad Q_* \ a.s.\\
&=& \int_0^{t} (1-\s(s))\beta\left( \int\a \s(s) \Phi_s dQ_* \right)
dM(s) ,\quad Q_* \ a.s.
\end{eqnarray*}
We now take the $L^2$-norm in both sides. For all $t\in [0,T]$ we
have
\begin{eqnarray*}
||\Phi_t||^2_{L^2(Q_*)}&=&\left\|\int_0^{t}(1-\s(s))\beta   \left(
\int\a \s(s) \Phi_s dQ_* \right
)dM(s)\ \right\|^2_{L^2(Q_*)}=\\
&=& E^{Q_*}\left[\int_0^{t}(1-\s(t))\beta^2  \left(  \int
\a\s(s)\Phi_s dQ_* \right)^2 e^{-\g+\b m_{Q_*}(s)} ds  \right].
\end{eqnarray*}
Notice that $ \left(  \int \a\s(s)\Phi_s  dQ_* \right)^2 \leq
\|\a\|^2\left( \int \Phi_s  dQ_* \right)^2 \leq\|\a\|^2 \int
\Phi_s^2 dQ_* \leq\|\a\|^2\int \Phi_t^2 dQ_* =\|\a\|^2
\|\Phi_t\|_{L^2(Q_*)}^2$, where $t\geq s$. The first inequality
follows since $\s\in\{0;1\}$;
 the second one is trivial and the latter one is due to the fact that $(\Phi_s^2)_s$
 is a submartingale and thus its expected value is an increasing function of time. Then
\begin{eqnarray*}
||\Phi_t||^2_{L^2(Q_*)}&\leq& \|\a\|^2E^{Q_*}
\left[\int_0^{t}(1-\s(t))\beta^2 \|\Phi_t\|_{L^2(Q_*)}^2 e^{-\g+\b m
_{Q_*}(s)} ds  \right]  \leq
  \ t \ \e^{-1}\    || \Phi_t||^2_{L^2(Q_*)}
\end{eqnarray*}
where $0<\e <\infty$ is a constant such that $\|\a\|^2\|\beta\|^2
E^{Q_*}[e^{-\g+\b m_{Q_*}(s)}]\leq \e^{-1} $.
As a consequence, $\Phi _s=0$, $\ Q_*\ a.s.$ for $s\in[0,\e)$.\\
This argument can be iterated defining $\Phi_t^{(2)}:=\Phi_{t+\e}$.
The same argument shows that    $\Phi_s^{(2)}=0$, $\ Q_*\ a.s.$ for
$s\in[0,\e)$; hence $\Phi_t=0$, $\ Q_*\ a.s.$ for $s\in[0,2\e)$.
Eventually we extend the statement to $s\in [0,T]$. Being $\Phi
_T=\Phi$, we would have $\hat\Phi=0$ and this gives a contradiction.
Hence the thesis follows.
 \epr

\noindent \emph{\underline{Step 3:}} Consider
$\lambda_1,\lambda_2\in B'$. Since $\lambda_i\circ T$, for $i=1,2$,
are in the topological dual of $\mathcal M$, there exist
$\Phi_1,\Phi_2\in\mathcal C_b$ such that $\lambda_i\circ T(Q)=\int
\Phi_i dQ$. We define
$$Cov_{p_*}(\lambda_1,\lambda_2)=\int\lambda_1(z)\lambda_2(z) p_*(dz)\quad \mbox{and}\quad \tilde \lambda_i=\int z \lambda_i (z) p_*(dz); \ i=1,2$$
where we recall that $p_*$, defined in $(B.4)$ of Theorem
\ref{TeoB}, is the centered version of the law of
$T(\delta_{\{\s[0,T],\l\}})$ induced by $Q_*$. Then the following
result holds true \vskip .5cm \bl{L.A.3.2}$\phantom{1}$
\begin{itemize}
\item[i)]$$Cov_{p_*}(\lambda_1,\lambda_2)=Cov_{Q_*}(\Phi_1,\Phi_2);$$
$$D^2\Psi(z_*)[\tilde \lambda_1,\tilde \lambda_2]=D^2F(Q_*)[\hat\Phi_1,\hat\Phi_2].$$
\item[ii)] For   $\lambda_i, \ i=1,2$  we have
$$\mathcal  Cov_{p_*}(\lambda_i,\lambda_i)-D^2\Psi(z_*)[\tilde \lambda_i,\tilde \lambda_i]>0.$$\end{itemize}
\el \Proof Point $(i)$. By the definition of $p_*$ and $\lambda_i,\
i=1,2$ we see that
$$Cov_{p_*}(\lambda_1,\lambda_2)=\int [T_{6+1} (\delta_{\{\s[0,T],\l\}})-T_{6+1} (Q_*)]
[T_{6+2}(\delta_{\{\s[0,T],\l\}})-T_{6+2}(Q_*)] dQ_* =$$
$$=\int [\Phi_1-\Phi_1^*][\Phi_2-\Phi_2^*] dQ_* ,$$
where we have used the fact that $\lambda_i \circ T (Q)=T_{6+i}(Q)=\int \Phi_i dQ$.\\
Concerning the second statement, we first prove the following claim.
\be{CC} \tilde\lambda_i=T(\hat\Phi_i)\ ;\ \   i=1,2.\ee To show the
validity of (\ref{CC}), we use the following two facts:
\begin{displaymath}\begin{array}{lcl}
\tilde\lambda_i&=&E^{Q_*}\left\{[T(\delta_{\{\s[0,T],\l\}})-T(Q_*)][\Phi_i(\s[0,T],\l)-\Phi_i^*]\right\};\\
\hat\Phi_i&=&E^{Q_*}\left\{[\delta_{\{\s[0,T],\l\}} - Q_*]
[\Phi_i(\s[0,T],\l)-\Phi_i^*]\right\}.\end{array}\end{displaymath}
The former follows by definition of $p_*$, $\lambda$ and $T_{6+i}
(Q)$, whereas the latter
is a consequence of the definition of $\hat\Phi$ given in (\ref{Hp}).\\
(\ref{CC}) is a consequence of the fact that $T$ is both linear and continuous,
hence we are allowed to interchange the $T$ operator with the expectation.\\
Having proved (\ref{CC}), we compute the second order Fr\'echet
derivatives on the function $\Psi$ as follows.
\be{D2Psi}D^2\Psi(z_*)[\tilde \lambda_1,\tilde \lambda_2]=
\lim_{k\to 0}\frac{D\Psi(z_*+k \tilde\lambda_2)[\tilde
\lambda_1]-D\Psi(z_*)[\tilde\lambda_1]}{k}.\ee Notice that, by the
linearity of $T$  and by (\ref{CC}), we have that
 $$z_*+k \tilde\lambda_2=T(Q_*+k \hat\Phi_2) \ ,\qquad z_*=T(Q_*).$$
 Thus
 $$\lim_{k\to 0}\frac{D\Psi(z_*+k \tilde\lambda_2)[\tilde \lambda_1]-D\Psi(z_*)[\tilde\lambda_1]}{k}=
 \lim_{k\to 0}\frac{D\Psi(z_*+k\tilde\lambda_2)[\lambda_1]-D\Psi(z_*)[\tilde\lambda_1]}{k
 }.$$
 We now claim that
 \be{marco}
 \lim_{k\to 0}\frac{D\Psi(z_*+k\tilde\lambda_2)[\lambda_1]-D\Psi(z_*)[\tilde\lambda_1]}{k}=
  \lim_{k\to
 0}\frac{DF(Q_*+ k \hat\Phi_2)[\hat\Phi_1]-DF(Q_*)[\hat\Phi_1]}{k}.\ee
 By (\ref{Fclaim2}) we see that
  $DF_2(Q_*)[\cdot]=0$ since $Q_*\ll (W\otimes \eta)$. Moreover both $DF_2(\hat\Phi_i)[\cdot]=0$ and $DF_2(Q_*+k  \hat\Phi_i)[\cdot]=0$
 since $\hat\Phi_i$ is absolutely
 continuous with respect to $Q_*$. This proves (\ref{marco}).
 Finally we use the fact that $F$ is Fr\'echet differentiable
 $$\lim_{k\to
 0}\frac{DF(Q_*+ k \hat\Phi_2)[\hat\Phi_1]-DF(Q_*)[\hat\Phi_1]}{k}=D^2F(Q_*)[\hat\Phi_1,\hat\Phi_2].$$
We have thus proved that $D^2\Psi(z_*)[\tilde \lambda_1,\tilde \lambda_2]=D^2F(Q_*)[\hat\Phi_1,\hat\Phi_2]$.\\

Concerning point $(ii)$, we notice that
\be{C=C}Cov_{p_*}(\lambda_i,\lambda_i)-D^2\Psi(z_*)[\tilde\lambda_i,\tilde\lambda_i]=Cov_{Q_*}(\Phi_i,\Phi_i)
-D^2F(Q_*)[\hat\Phi_i,\hat\Phi_i]=\mathcal H(\Phi_i,\Phi_i),\ee
where $\mathcal H$ has been defined in Equation (\ref{D3A}). Hence
by Proposition \ref{Prop1}  the positivity condition is ensured
and the thesis follows.\fine

By virtue of Lemma \ref{L.A.3.2}, for any $\lambda\in B'$ such
that $\tilde \lambda \ne 0$,
 (\ref{varL}) holds true. As a consequence, assumption $(B.4)$ is ensured and thus Theorem \ref{t3} is proved.
\fine

\end{document}